\documentclass[11pt]{article}
\usepackage{amssymb}
\usepackage{amsfonts}
\usepackage{amsmath}

\newtheorem{theorem}{Theorem}

\newtheorem{claim}{Claim}

\newtheorem{corollary}{Corollary}

\newtheorem{lemma}{Lemma}

\newtheorem{proposition}{Proposition}
\newtheorem{remark}{Remark}

\begin{document}

\title{Equal Treatment of Equals and Efficiency in Probabilistic Assignments%
\thanks{%
An earlier version of this paper was presented at the 2026 Spring Meeting of
the Japanese Economic Association at Osaka Metropolitan University. The
author thanks Minoru Kitahara and Kenzo Imamura for their insightful
comments and suggestions. This work was supported by JSPS KAKENHI Grant
Number 25K05004.}}
\author{Yasunori Okumura\thanks{%
Department of Logistics and Information Engineering, Tokyo University of
Marine Science and Technology (TUMSAT), 2-1-6 Etchujima, Koto-ku, Tokyo
135-8533, Japan, Phone: +81-3-5245-7300, Fax: +81-3-5245-7300, E-mail:
yokumu0@kaiyodai.ac.jp}}
\maketitle

\begin{center}
\textbf{Abstract}
\end{center}

This paper studies general multi-unit assignment problems with indivisible
objects, focusing on equal treatment of equals (ETE). We adopt a
lottery-based framework because the standard stochastic-matrix approach does
not capture the joint structure of multi-unit assignments, and generalize
ETE to accommodate a broad range of constraints and applications. We
introduce the ETE reassignment procedure, which transforms any assignment
into one satisfying ETE, and examine whether it preserves ex-post
efficiency, ordinal efficiency, and rank-minimizing efficiency. We show that
it preserves ex-post efficiency but not necessarily ordinal efficiency.
However, the ETE reassignment of a rank-minimizing assignment remains
rank-minimizing and hence ordinally efficient, establishing the existence of
an assignment satisfying both ETE and ordinal efficiency. Finally, under
general upper-bound constraints, we provide a computationally efficient
construction of such an assignment by combining serial dictatorship,
appropriately chosen priority lists, and ETE reassignment.


\newpage

\section{Introduction}

This paper studies general multi-unit assignment problems with indivisible
objects, focusing on the fairness requirement known as equal treatment of
equals (ETE), which requires that agents with identical relevant
characteristics receive equal treatment. Rooted in Aristotle's Nicomachean
Ethics (Book V), ETE represents an intuitive notion of fairness. A large
body of empirical and experimental research suggests that individuals
dislike unfair outcomes and are often willing to bear costs to avoid them;
see, for example, Fehr and Charness (2025) for a survey. This suggests that,
even when efficiency is a central concern, allocation rules should satisfy
at least a minimal fairness requirement such as ETE.

However, in environments with indivisible object types, strict enforcement
of ETE may lead to inefficiencies when the number of equals demanding an
object type exceeds its supply. To address this issue, the literature,
including Hylland and Zeckhauser (1979), Bogomolnaia and Moulin (2001),
Budish et al. (2013), Erdil (2014), and Basteck and Ehlers (2025), considers
probabilistic assignments. Under such assignments, fairness is preserved by
requiring equals to receive identical probability distributions over object
types rather than identical deterministic outcomes.

In this literature, agents with identical preference orderings are typically
regarded as equals. This definition raises several practical difficulties.
First, it is incompatible with policy goals such as affirmative action,
because it precludes giving preferential treatment to disadvantaged agents
who report the same preferences as advantaged ones. Second, if agents who
are asymmetric in terms of feasibility constraints are nevertheless regarded
as equals, achieving efficiency may become difficult. Accordingly, we adopt
an extended notion of ETE that avoids these problems.

In this paper, we investigate whether an assignment that satisfies (the
extended) ETE can also satisfy additional desirable properties, and, if so,
how such an assignment can be constructed. To obtain an ETE assignment, we
use the following simple procedure. First, we arbitrarily fix a pure
assignment, in which each agent receives an individual assignment with
probability one. Second, for each group of equals, we pool the individual
assignments they receive under this pure assignment and redistribute them
uniformly within the group, so that each agent receives each of these
assignments with equal probability. This procedure is called the ETE
reassignment of an assignment. It can be naturally extended to cases where
the original assignment is not pure but probabilistic. In general settings,
the ETE reassignment yields an assignment that satisfies ETE.

A distinctive feature of this approach is that it treats a probabilistic
assignment as a lottery over pure assignments, rather than as a stochastic
matrix of assignment probabilities. Although the stochastic-matrix
formulation is often sufficient in unit-demand environments, it is generally
not rich enough for multi-unit assignment problems, because it records only
the probability with which each agent receives each object type and
abstracts from how objects are combined into bundles.\footnote{%
Nguyen et al. (2016) also emphasize the importance of representing
probabilistic assignments at the bundle level in environments with
multi-unit demands and complementarities.} This bundle-level structure is
essential for capturing complementarity and substitutability among objects.
By contrast, the lottery formulation preserves this structure.

The key question is whether the properties satisfied by the original
assignment are preserved under its ETE reassignment. If they are preserved,
then obtaining the ETE reassignment itself is straightforward. Hence, for
our purposes, it suffices to construct an original assignment that satisfies
the target properties. The difficulty of this task depends on how general
the feasibility constraints are. We first examine the most basic
environment, which we call the simple-constraint case: agents have unit
demand and feasibility is described by simple capacity constraints. We then
consider more general feasibility constraints.

In this study, we focus on three efficiency notions from the existing
literature as the properties of interest. First, we consider ex-post
efficiency (hereafter EE), which requires that every pure assignment
realized with positive probability be (Pareto) efficient. It is immediate
that the ETE reassignment of an EE assignment remains EE. Second, we focus
on ordinal efficiency (OE), which is stronger than EE. An assignment is OE
if it is not first-order stochastically dominated by any other assignment
for every agent, with strict dominance for at least one agent.
Unfortunately, in general, the ETE reassignment of an OE assignment need not
be OE. Third, we consider a stronger notion of efficiency, known as
rank-minimizing efficiency (hereafter RE). An RE assignment always exists
and is OE. Moreover, since the ETE reassignment of an RE assignment remains
RE, this establishes the existence of an assignment satisfying both ETE and
RE (and hence OE).

However, except in the simple-constraint case, no computationally efficient
method is known for finding an RE assignment. Thus, although the ETE
reassignment procedure itself is easy to implement, the existence result
does not generally provide a computationally efficient construction. In the
simple-constraint case, by contrast, an RE assignment can be computed
efficiently using an existing method. Combining this method with the ETE
reassignment procedure yields a computationally efficient construction of an
assignment satisfying both ETE and RE, and hence OE. For more general
settings, we instead develop a computationally efficient method for
constructing an assignment satisfying both ETE and OE without first
computing an RE assignment.

We consider the case where feasible assignments are restricted by general
upper bounds. In this setting, the serial dictatorship rule with an
arbitrary priority list yields an OE assignment; however, the ETE
reassignment of this assignment may fail to be OE. In contrast, if the
priority list satisfies a property called consecutive equals, the serial
dictatorship rule yields an OE assignment whose ETE reassignment is also OE.

\section{Related Literature}

We consider a general multi-unit assignment model, which encompasses, for
example, course allocation problems; see S\"{o}nmez and \"{U}nver (2010) and
Budish and Cantillon (2012). Nguyen et al. (2016) provide several additional
examples of real-world allocation problems that inherently involve
multi-unit demands.

Kojima (2009) studies probabilistic assignments in a multi-unit assignment
model, and Balbuzanov (2022) generalizes Kojima's framework by incorporating
general feasibility constraints, motivated by the various restrictions that
arise in practical course allocation problems, such as scheduling
constraints. In the models of Kojima (2009) and Balbuzanov (2022), agents
are assumed to have restricted preferences: only the expected number of
copies received from each type matters. By contrast, we allow agents to have
general preferences over bundles and represent probabilistic assignments as
lotteries over pure assignments.

Kesten et al. (2017) also represent probabilistic assignments as lotteries
over pure assignments, emphasizing their practical implementability and the
direct analysis of ex-post properties. Unlike their unit-demand setting,
however, our model allows agents to receive multiple objects. Consequently,
the lottery formulation plays an additional role in our setting: it
preserves the joint structure of multi-unit assignments, which is generally
lost under the stochastic-matrix representation.

We focus on a relatively weak fairness notion, ETE. This concept has been
studied extensively across a wide range of allocation problems (Varian,
1974; Moulin, 2004; Thomson, 2011; Yokote et al., 2019). In single-unit
assignment problems, including one-to-many matching environments, ETE has
been analyzed by Bogomolnaia and Moulin (2001). ETE is weaker than several
other fairness notions, such as envy-freeness discussed by many studies
including Hylland and Zeckhauser (1979). These stronger notions typically
impose requirements that go beyond the na\"{\i}ve definition of ETE, which
requires only that agents with identical preferences receive identical
assignments. As a result, such fairness notions may be incompatible with
certain policy objectives, including affirmative action, which intentionally
prescribes differential treatment among agents who may share the same
preferences.

We consider three efficiency notions: EE, OE, and RE. Among them, RE is the
strongest, followed by OE, and EE is the weakest. OE is a widely used
concept in the probabilistic assignment literature, including Bogomolnaia
and Moulin (2001) and Budish et al. (2013). Recently, several studies, such
as Featherstone (2020), have focused on RE, as many real-world matching
authorities take the rank positions of assignments into account. Since Feizi
(2024) shows that several fairness notions other than ETE are incompatible
with RE, it is appropriate to adopt ETE as the fairness notion along with RE.

We study the case in which feasible assignments are subject to general upper
bound constraints. Our general constraint structure encompasses, as special
cases, the regional caps discussed by Kamada and Kojima (2015), as well as
the \textit{object-type-specific} general upper bound constraints considered
by Okumura (2019) and Kamada and Kojima (2024). While Imamura and Kawase
(2025) consider this general constraint structure, their model is limited to
the unit-demand setting.\footnote{%
Kamada and Kojima (2024) and Imamura and Kawase (2025) illustrate that
general upper-bound constraints are applicable to a wide range of real-world
settings, such as the refugee matching problems (Andersson and Ehlers,
2020), the day-care matching problems (Okumura, 2019), and the controlled
school choice problems (Abdulkadiro\u{g}lu and S\"{o}nmez, 2003).} In the
multi-unit demand case with general upper bound constraints, we introduce a
method for deriving an assignment that satisfies both OE and ETE.

We compare our method, which applies the ETE reassignment, with several
existing methods proposed in the literature. To begin with, it is worth
noting that these existing methods were originally designed to satisfy the
standard notion of ETE. As emphasized above, mechanisms that provide
preferential treatment to disadvantaged agents through affirmative action do
not satisfy this standard notion. Moreover, the standard notion is not
compatible with several efficiency concepts when general constraint
structures are allowed. Therefore, we extend the concept of ETE and develop
a method for deriving assignments that satisfy this extended notion.

First, Nikzad (2022), Ortega and Klein (2023), Troyan (2024), and Okumura
(2026a) study the uniform rank-minimizing mechanism, which satisfies both RE
and ETE in the simple-constraint case. However, as noted by Troyan (2024,
footnote 11), this mechanism suffers from a computational drawback: finding
all rank-minimizing assignments is computationally infeasible. In contrast,
our method is computationally efficient for constructing an assignment that
satisfies both RE and ETE in the simple-constraint case.

Second, the random serial dictatorship rule is discussed in several previous
studies, such as Bogomolnaia and Moulin (2001), and yields assignments
satisfying both EE and ETE. However, as shown by Bogomolnaia and Moulin
(2001), the outcome of this rule may fail to satisfy OE. In contrast, we
provide a method that yields assignments satisfying both OE and ETE under a
general constraint structure.

Third, Bogomolnaia and Moulin (2001) introduce the probabilistic serial
mechanism, which satisfies both OE and envy-freeness, and hence ETE. Budish
et al. (2013) generalize this mechanism and show that it yields assignments
satisfying both OE and envy-freeness under general constraint structures.
However, in their extension of the probabilistic serial mechanism, Budish et
al. (2013) restrict their attention to single-unit demand settings.

Balbuzanov (2022) also proposes a generalized probabilistic serial mechanism
that applies under more general constraints with multi-unit demand and shows
that its outcome satisfies both OE and ETE. Nevertheless, for this mechanism
to attain efficiency, it is necessary that agents have specific preference
structures, as explained above.

Nguyen et al. (2016) also generalize the probabilistic serial mechanism and
show that it yields assignments satisfying both OE and envy-freeness even
when agents' ordinal preferences exhibit a limited degree of
complementarities, provided that bounded violations of simple capacity
constraints are allowed. In contrast, our method does not rely on such
approximate feasibility: although we adopt a weaker fairness requirement, we
impose feasibility strictly and remain applicable to more general preference
profiles.

It should be noted that the mechanism that naively applies the results of
this study is vulnerable to strategic manipulation. For comparison, as shown
by Bogomolnaia and Moulin (2001) and Budish et al. (2013), the random serial
dictatorship rule satisfies strategy-proofness in a strict sense, whereas
the probabilistic serial mechanism satisfies a weaker form of this property.%
\footnote{%
Basteck and Ehlers (2025) introduce another mechanism that satisfies EE,
ETE, and strategy-proofness.} In contrast, the mechanism that naively
applies our results fails to satisfy strategy-proofness even in a weak
sense; that is, under truth-telling, an agent's outcome can be first-order
stochastically dominated by that under a manipulation.

Even in the existing literature on multi-unit demand models, achieving
strategy-proofness has been shown to be difficult. In particular, Balbuzanov
(2022) shows that their generalized probabilistic serial mechanism fails to
satisfy weak strategy-proofness in the considered setting. Moreover,
Kornbluth et al. (2025) establish that the generalized probabilistic serial
mechanism proposed by Nguyen et al. (2016) likewise does not satisfy weak
strategy-proofness. These results suggest that, in general multi-unit demand
models, imposing both fairness and strategy-proofness may be difficult, even
when strategy-proofness is required only in a weak sense. Therefore, in this
paper, we abstract from strategy-proofness and focus instead on mechanisms
that are efficient and fair.

Finally, the ETE reassignment introduced in this paper is also applicable to
the probabilistic school choice model of Kesten and \"{U}nver (2015).
Indeed, Okumura (2026b) shows that the ETE reassignment of a deterministic
stable matching remains stable in the sense of ex ante stability.

\section{Model}

Let $A$ and $O$ be finite sets of agents and object types respectively. A 
\textbf{pure assignment} $y$ is a $\left\vert A\right\vert \times \left\vert
O\right\vert $ matrix, where $y_{ao}\in \mathbb{Z}_{+}$ represents the
number of copies of object type $o$ assigned to agent $a$. Moreover, let $%
y_{a}=\left( y_{ao}\right) _{o\in O}\in \mathbb{Z}_{+}^{\left\vert
O\right\vert }$ and $y_{o}=\left( y_{ao}\right) _{a\in A}\in \mathbb{Z}%
_{+}^{\left\vert A\right\vert }.$

Pure assignments are subject to constraints. A pure assignment that
satisfies these constraints is called a \textbf{feasible pure assignment}.
Let $Y$ be the set of all feasible pure assignments, which is assumed to be
a non-empty and finite set.

The set $Y$ may be subject to various types of constraints, including
physical capacity constraints and institutional constraints. In addition, if
one is interested only in assignments that satisfy certain desirable
properties, $Y$ can be defined as the set of all assignments satisfying
those properties. For instance, given a priority structure, if attention is
restricted to stable assignments, then $Y$ coincides with the set of all
stable assignments.

Let $X\subseteq \mathbb{Z}_{+}^{\left\vert O\right\vert }$ be the set of all
possible pure (individual) assignments for $a\in A$. We assume that $X$ is
finite. An integer vector $x=\left( x_{1},\cdots ,x_{\left\vert O\right\vert
}\right) \in \mathbb{Z}_{+}^{\left\vert O\right\vert }$ is said to be a 
\textbf{feasible pure assignment} \textbf{for} $a$ if $x=y_{a}$ for some $%
y\in Y$. Note that $X$ may include some infeasible pure assignment for some
agents, since the feasibility constraints encoded in $Y$ may be complex.
Specifically, if $x_{o}=1$ and $x_{o^{\prime }}=0$ for all $o^{\prime }\in
O\setminus \left\{ o\right\} $, we simply write $x=o$. Moreover, $\mathbf{0}%
=\left( 0,\cdots ,0\right) $ denotes the zero vector.

Likewise, an integer vector $z=\left( z_{1},\cdots ,z_{\left\vert
A\right\vert }\right) \in \mathbb{Z}_{+}^{\left\vert A\right\vert }$ is said
to be a \textbf{feasible pure assignment} \textbf{for} $o$ if $z=y_{o}$ for
some $y\in Y$.

We introduce standard constraints in the single-unit assignment model.
First, we say that $Y$ satisfies\textbf{\ (single) unit demand }if\textbf{\ }%
$y\in Y$ implies 
\begin{equation*}
\sum\limits_{o\in O}y_{ao}=1\text{ for all }a\in A.
\end{equation*}%
Next, let $q_{o}\in \mathbb{Z}_{++}$ be the capacity (i.e., the number of
copies) of $o\in O$. We say that $Y$ satisfies the\textbf{\ simple capacity
constraint} if $y\in Y$ implies 
\begin{equation*}
\sum\limits_{a\in A}y_{ao}\leq q_{o}\text{ for all }o\in O\text{.}
\end{equation*}%
If $Y$ satisfies unit demand and simple capacity constraints, then we say
that $Y$ falls under \textbf{the simple-constraint case}.

Let $\succsim _{a}$ be a preference relation of $a$ over $X$, where $x\succ
_{a}x^{\prime }$ means that $a$ prefers $x$ to $x^{\prime }$ and $x\succsim
_{a}x^{\prime }$ means $x\succ _{a}x^{\prime }$ or $x=x^{\prime }$. Note
that $y_{a}\succ _{a}y_{a}^{\prime }$ indicates that $a$ prefers the pure
assignment $y$ to $y^{\prime }$. A feasible pure assignment $y\in Y$ is 
\textbf{efficient} if there is no feasible pure assignment $y^{\prime }\in Y$
that Pareto dominates $y$; that is, $y_{a}^{\prime }\succ _{a}y_{a}$ for
some $a\in A$ and $y_{b}^{\prime }\succsim _{b}y_{b}$ for all $b\in A$.

Let $A_{1},\cdots ,A_{N}$ be a partition of $A$, where $N\leq \left\vert
A\right\vert $, and for any $n=1,\cdots ,N$, $a,b\in A_{n}$ if and only if $%
a $ and $b$ are \textit{equals}. We refer to each $A_{n}$ as a\textbf{\
group of equals}.

Throughout this paper, we impose the following two assumptions on the groups
of equals.

\begin{description}
\item[Assumption 1] For each $n=1,\cdots ,N$ and any $a,b\in A_{n}$,\textit{%
\ }$\succsim _{a}=\succsim _{b}$.
\end{description}

Assumption 1 means that two agents are regarded as equals \textit{only if}
their preference orders are identical. Note that $a$ and $b$ may belong to
different groups of equals even if $\succsim _{a}=\succsim _{b}$. For
example, if affirmative action permits giving priority to racial minority
students in assignments, then two students with different races are no
longer considered \textquotedblleft equals\textquotedblright .

\begin{description}
\item[Assumption 2] \textit{For any two agents (equals) }$a,b\in A_{n}$%
\textit{\ for some }$n=1,\cdots ,N$\textit{, let }$y$\textit{\ and }$%
y^{\prime }$\textit{\ be two pure assignments such that }$%
y_{ao}=y_{bo}^{\prime }$\textit{\ and }$y_{bo}=y_{ao}^{\prime }$\textit{\
for all }$o\in O,$\textit{\ and }$y_{co}=y_{co}^{\prime }$\textit{\ for all }%
$c\in A\setminus \left\{ a,b\right\} $ and \textit{all }$o\in O$\textit{.
Then, }$y\in Y$\textit{\ implies }$y^{\prime }\in Y$\textit{.}
\end{description}

Assumption 2 means that if two agents are equals, then the feasibility of an
assignment is unchanged even if their assignments are exchanged. That is, in
this paper, equals are assumed to be equals also with respect to the
constraints. This assumption is also adopted in Balbuzanov (2022).

As an illustrative example, we consider the Japanese day-care matching
market (Okumura, 2019). An assignment may become infeasible if an older
child is replaced by an infant, due to differences in staffing requirements
and space allocation. Accordingly, in such markets, two agents of different
ages may belong to different groups of equals even if their preference
orders are identical.

By Assumptions 1 and 2, agents are regarded as equals \textit{only if }they
have identical preferences and face identical constraints. However, in
contrast to the definition of Balbuzanov (2022), we allow that agents with
identical preferences and constraints are not equals.

Section 6 discusses which characteristics should be regarded as available,
paying particular attention to policy considerations and the potential
incentive problems that may arise when assignment decisions rely on personal
characteristics.

Next, we consider a (\textbf{probabilistic) assignment}, which is
represented by a lottery over $Y$ denoted by $\sigma :Y\rightarrow \left[ 0,1%
\right] $, such that 
\begin{equation*}
\sum\limits_{y\in Y}\sigma \left( y\right) =1\text{,}
\end{equation*}%
where $\sigma \left( y\right) $ represents the probability that pure
assignment $y$ is realized.\footnote{%
Note that Kesten et al. (2017) also represent assignments as probability
distributions over pure assignments and point out several advantages of this
approach. In their terminology, the mechanism we consider is a lottery
mechanism.} Let $\Sigma $ be the set of all possible lotteries over $Y$.
That is, we focus on assignments that are implementable as a lottery over
feasible pure assignments. Moreover, let 
\begin{equation*}
\mathrm{Supp}(\sigma )=\left\{ y\in Y\text{ }\left\vert \text{ }\sigma
\left( y\right) >0\right. \right\}
\end{equation*}%
be the support of $\sigma $. Specifically, for each pure assignment $y$, let 
$\sigma ^{y}$ denote the deterministic assignment over pure assignments that
assigns probability one to $y$; that is, $\sigma ^{y}\left( y\right) =1$.

An assignment $\sigma $ is said to be \textbf{ex-post efficient (EE)} if any 
$y\in \mathrm{Supp}(\sigma )$ is efficient. In the subsequent section, we
consider two other efficiency notions, both of which are stronger than
ex-post efficiency.

Let $\mathbf{x}$ be a random variable where $\mathbf{x}=x\in X$ with
probability $\Pr \left( x;\mathbf{x}\right) $. Let $\mathbf{x}\left( \sigma
\right) _{a}$ be a random variable representing the individual assignment of 
$a$ under $\sigma $; that is, 
\begin{equation*}
\Pr \left( x;\mathbf{x}\left( \sigma \right) _{a}\right) =\sum\limits_{y:%
\text{ }y_{a}=x}\sigma \left( y\right) \text{.}
\end{equation*}

We consider stochastic dominance relations of these random variables. For $%
a\in A,$ let $\left\{ x^{1},x^{2},\cdots ,x^{\left\vert X\right\vert
}\right\} =X$ be such that $x^{i}\succ _{a}x^{i+1}$ for all $i=1,2,\cdots
,\left\vert X\right\vert -1$. Let%
\begin{equation*}
F_{a}\left( x,\mathbf{x}\right) =\sum\limits_{i=i^{\prime }+1}^{\left\vert
X\right\vert }\Pr \left( x^{i};\mathbf{x}\right) ,\text{ where }%
x=x^{i^{\prime }}\text{ }
\end{equation*}%
represents the probability that the pure assignment of $a$ is \textit{less
preferable} than $x$ under $\mathbf{x}$. Let 
\begin{equation*}
\bar{F}_{a}\left( x,\mathbf{x}\right) =1-F_{a}\left( x,\mathbf{x}\right) ,
\end{equation*}%
which represents the probability that the pure assignment of $a$ is more
preferable than or is equal to $x$ under $\mathbf{x}$. Then, a random
variable $\mathbf{x}$ is \textbf{first-order stochastically dominated }by $%
\mathbf{x}^{\prime }$ for agent $a$\textbf{\ }if 
\begin{equation*}
\bar{F}_{a}\left( x^{i},\mathbf{x}^{\prime }\right) \geq \bar{F}_{a}\left(
x^{i},\mathbf{x}\right)
\end{equation*}%
for all $i=1,2,\cdots ,\left\vert X\right\vert $. Moreover, $\mathbf{x}$ is 
\textbf{strictly} \textbf{first-order stochastically dominated }by $\mathbf{x%
}^{\prime }$ for agent $a$\textbf{\ }if $\mathbf{x}$ is first-order
stochastically dominated\textbf{\ }by $\mathbf{x}^{\prime }$ for agent $a$%
\textbf{\ }and $\mathbf{x\neq x}^{\prime }$.

These concepts differ from those in Kojima (2009) and Balbuzanov (2022). In
their models, probabilistic assignments are represented as matrices of
expected numbers of copies of each object type, because agents' preferences
depend only on the expected number of copies of each object type.\footnote{%
In fact, Kojima (2009) explicitly assumes that preferences over sets of
objects are additively separable and acknowledges in footnote 7 that this
assumption is restrictive. More generally, the same limitation applies to
analyses of non-unit-demand environments based on stochastic matrices,
because such matrices record only marginal assignment probabilities and
therefore abstract from complementarities, substitutabilities, and other
joint features of individual assignments. See also Nguyen et al. (2016), who
emphasize the importance of complementarities in multi-unit assignment
problems.} In order to ensure that probabilistic assignments can be
represented in this way, they generalize the Birkhoff--von Neumann theorem.
By contrast, our approach does not rely on such a representation.

To illustrate how restrictive it is to focus only on the expected number of
each object type, consider the following example.

\subsubsection*{Example 1}

Let 
\begin{eqnarray*}
y_{a} &=&\left( y_{ao_{1}},y_{ao_{2}},y_{ao_{3}}\right) =(1,1,0), \\
y_{a}^{\prime } &=&(0,1,1),y_{a}^{\prime \prime }=(0,0,1),y_{a}^{\prime
\prime \prime }=(1,0,0).
\end{eqnarray*}%
Let $\sigma $ and $\sigma ^{\prime }$ be such that $\sigma \left( y\right)
=\sigma \left( y^{\prime \prime }\right) =0.5$ and $\sigma ^{\prime }\left(
y^{\prime }\right) =\sigma ^{\prime }\left( y^{\prime \prime \prime }\right)
=0.5$. Then, 
\begin{eqnarray*}
\sigma \left( y\right) \times (1,1,0)+\sigma \left( y^{\prime \prime
}\right) \times (0,0,1) &=& \\
\sigma ^{\prime }\left( y^{\prime }\right) \times (0,1,1)+\sigma ^{\prime
}\left( y^{\prime \prime \prime }\right) \times (1,0,0) &=&\left(
0.5,0.5,0.5\right) ;
\end{eqnarray*}%
that is, the expected number of copies of each object type assigned to $a$
is identical between two assignments $\sigma $ and $\sigma ^{\prime }$.

First, suppose that $o_{1}$ and $o_{2}$ are substitutes and $o_{3}$ is an
independent good. Let the preference ordering be given by $%
x^{1}=y_{a}^{\prime },$ $x^{2}=y_{a},$ $x^{3}=y_{a}^{\prime \prime \prime }$
and $x^{4}=y_{a}^{\prime \prime }$ ($x^{i}\succ _{a}x^{i+1}$ for all $%
i=1,2,3 $). By our definition, in the first case, $\mathbf{x}\left( \sigma
\right) _{a}$ is strictly first-order stochastically dominated by $\mathbf{x}%
\left( \sigma ^{\prime }\right) _{a}$\textbf{\ }for agent $a$. Second,
suppose that $o_{1}$ and $o_{2}$ are complements and $o_{3}$ is an
independent good. In this case, we let the preference ordering be given by $%
x^{1}=y_{a},$ $x^{2}=y_{a}^{\prime },$ $x^{3}=y_{a}^{\prime \prime }$ and $%
x^{4}=y_{a}^{\prime \prime \prime }$. In this case, $\mathbf{x}\left( \sigma
^{\prime }\right) _{a}$ is strictly first-order stochastically dominated by $%
\mathbf{x}\left( \sigma \right) _{a}$\textbf{\ }for agent $a$. Therefore,
evaluating assignments solely based on the expected number of copies of each
object type assigned in multi-unit demand models is restrictive.\newline

We say that an assignment $\sigma $ satisfies \textbf{equal treatment of
equals} (hereafter \textbf{ETE}) if for any two agents $a$ and $b$ belonging
to the same group of equals, $\mathbf{x}\left( \sigma \right) _{a}=\mathbf{x}%
\left( \sigma \right) _{b}$ meaning that $\mathbf{x}\left( \sigma \right)
_{a}$ and $\mathbf{x}\left( \sigma \right) _{b}$ have the same distribution.

We now define the ETE reassignment, a procedure for deriving an assignment
that satisfies ETE from a given initial assignment. We define a mapping $%
\epsilon :\Sigma \rightarrow \Sigma $ as follows. For each $\sigma \in
\Sigma $, we call $\epsilon \left( \sigma \right) $ the \textbf{ETE
reassignment }of $\sigma $.

First, we consider $\epsilon \left( \sigma ^{y}\right) $ which is the ETE
reassignment of a deterministic assignment. Let $A_{n}=\left\{ a_{i},\cdots
,a_{i+j}\right\} $ be agents who are equals, where $j\geq 1$. Then, their
individual assignments $y_{a_{i}},\cdots ,y_{a_{i+j}}$ may differ from one
another. In $\epsilon \left( \sigma ^{y}\right) $, these individual
assignments are pooled and then redistributed among them, so that each agent
receives each of these assignments with equal probability $1/\left(
j+1\right) $.

Formally, let $\pi :A\rightarrow A$ be a bijection satisfying $\pi \left(
a\right) =b$ implies $a,b\in A_{n}$ for some $n=1,\cdots ,N$. Let $\pi
^{1},\pi ^{2},\cdots ,\pi ^{L}$ be distinct possible such bijections where 
\begin{equation*}
L=\left\vert A_{1}\right\vert !\times \cdots \times \left\vert
A_{N}\right\vert !.
\end{equation*}%
Moreover, let 
\begin{equation*}
L_{-n}=\left\vert A_{1}\right\vert !\times \cdots \times \left\vert
A_{n-1}\right\vert !\times \left\vert A_{n+1}\right\vert !\times \cdots
\times \left\vert A_{N}\right\vert !.
\end{equation*}%
Fix an arbitrary pure assignment $y\in Y$. We let for $l=1,\cdots ,L$, $%
y^{l} $ be such that $y_{\pi ^{l}(a)o}^{l}=y_{ao}$ for all $a\in A$ and $%
o\in O$. We let a multiset $Y_{D}\left( y\right) =\left\{ \left\{
y^{1},\cdots ,y^{L}\right\} \right\} $ and say that each element of $%
Y_{D}\left( y\right) $ is \textbf{derived from} $y$. Note that any $y$ is
derived from itself. Moreover, even if $l\neq l^{\prime },$ $%
y^{l}=y^{l^{\prime }}$ may be satisfied, because two agents in the same
group of equals may receive the same individual assignment under $y$.
Therefore, $Y_{D}\left( y\right) $ is a multiset. Furthermore, by Assumption
2, every assignment that is derived from any feasible assignment is also
feasible. Now, we let $\epsilon \left( \sigma ^{y}\right) $ be the
assignment such that 
\begin{equation*}
\epsilon \left( \sigma ^{y}\right) \left( y^{\prime }\right) =\frac{%
\left\vert \left\{ l=1,\cdots ,L\text{ }\left\vert \text{ }y^{l}=y^{\prime
}\right. \right\} \right\vert }{L}\text{.}
\end{equation*}

On the other hand, we consider a general assignment $\sigma $. For each $%
y^{\prime }\in Y$,

\begin{equation*}
\epsilon \left( \sigma \right) \left( y^{\prime }\right) =\sum\limits_{y\in
Y}\sigma \left( y\right) \times \epsilon \left( \sigma ^{y}\right) \left(
y^{\prime }\right) .
\end{equation*}

We introduce an example to understand the ETE reassignment procedure.

\subsubsection*{Example 2}

Let%
\begin{align*}
y& =\left( 
\begin{array}{ccccc}
y_{a_{1}o_{1}} & y_{a_{1}o_{2}} & y_{a_{1}o_{3}} & y_{a_{1}o_{4}} & 
y_{a_{1}o_{5}} \\ 
y_{a_{2}o_{1}} & y_{a_{2}o_{2}} & y_{a_{2}o_{3}} & y_{a_{2}o_{4}} & 
y_{a_{2}o_{5}} \\ 
y_{a_{3}o_{1}} & y_{a_{3}o_{2}} & y_{a_{3}o_{3}} & y_{a_{3}o_{4}} & 
y_{a_{3}o_{5}} \\ 
y_{a_{4}o_{1}} & y_{a_{4}o_{2}} & y_{a_{4}o_{3}} & y_{a_{4}o_{4}} & 
y_{a_{4}o_{5}} \\ 
y_{a_{5}o_{1}} & y_{a_{5}o_{2}} & y_{a_{5}o_{3}} & y_{a_{5}o_{4}} & 
y_{a_{5}o_{5}}%
\end{array}%
\right) =\left( 
\begin{array}{ccccc}
0 & 1 & 0 & 0 & 0 \\ 
1 & 0 & 0 & 0 & 0 \\ 
0 & 0 & 1 & 0 & 0 \\ 
0 & 0 & 0 & 1 & 0 \\ 
0 & 0 & 0 & 0 & 1%
\end{array}%
\right) , \\
y^{\prime }& =\left( 
\begin{array}{ccccc}
0 & 0 & 1 & 0 & 0 \\ 
0 & 0 & 0 & 1 & 0 \\ 
0 & 1 & 0 & 0 & 0 \\ 
1 & 0 & 0 & 0 & 0 \\ 
0 & 0 & 0 & 0 & 1%
\end{array}%
\right) .
\end{align*}%
We assume that $y$ and $y^{\prime }$ are feasible. Let $\sigma $ be such
that $\sigma \left( y\right) =1/3,$ and $\sigma \left( y^{\prime }\right)
=2/3$. Now, suppose that $A_{1}=\left\{ a_{1},a_{2}\right\} $, $%
A_{2}=\left\{ a_{3},a_{4}\right\} $ and $A_{3}=\left\{ a_{5}\right\} $. Let $%
y^{1},\cdots ,y^{4}$ and $y^{\prime 1},\cdots ,y^{\prime 4}$ be pure
assignments derived from $y$ and $y^{\prime }$ respectively, where%
\begin{align*}
y^{1}& =\left( 
\begin{array}{ccccc}
0 & 1 & 0 & 0 & 0 \\ 
1 & 0 & 0 & 0 & 0 \\ 
0 & 0 & 1 & 0 & 0 \\ 
0 & 0 & 0 & 1 & 0 \\ 
0 & 0 & 0 & 0 & 1%
\end{array}%
\right) ,y^{2}=\left( 
\begin{array}{ccccc}
1 & 0 & 0 & 0 & 0 \\ 
0 & 1 & 0 & 0 & 0 \\ 
0 & 0 & 1 & 0 & 0 \\ 
0 & 0 & 0 & 1 & 0 \\ 
0 & 0 & 0 & 0 & 1%
\end{array}%
\right) ,y^{3}=\left( 
\begin{array}{ccccc}
0 & 1 & 0 & 0 & 0 \\ 
1 & 0 & 0 & 0 & 0 \\ 
0 & 0 & 0 & 1 & 0 \\ 
0 & 0 & 1 & 0 & 0 \\ 
0 & 0 & 0 & 0 & 1%
\end{array}%
\right) \\
y^{4}& =\left( 
\begin{array}{ccccc}
1 & 0 & 0 & 0 & 0 \\ 
0 & 1 & 0 & 0 & 0 \\ 
0 & 0 & 0 & 1 & 0 \\ 
0 & 0 & 1 & 0 & 0 \\ 
0 & 0 & 0 & 0 & 1%
\end{array}%
\right) ,y^{\prime 1}=\left( 
\begin{array}{ccccc}
0 & 0 & 1 & 0 & 0 \\ 
0 & 0 & 0 & 1 & 0 \\ 
0 & 1 & 0 & 0 & 0 \\ 
1 & 0 & 0 & 0 & 0 \\ 
0 & 0 & 0 & 0 & 1%
\end{array}%
\right) ,y^{\prime 2}=\left( 
\begin{array}{ccccc}
0 & 0 & 0 & 1 & 0 \\ 
0 & 0 & 1 & 0 & 0 \\ 
0 & 1 & 0 & 0 & 0 \\ 
1 & 0 & 0 & 0 & 0 \\ 
0 & 0 & 0 & 0 & 1%
\end{array}%
\right) , \\
y^{\prime 3}& =\left( 
\begin{array}{ccccc}
0 & 0 & 1 & 0 & 0 \\ 
0 & 0 & 0 & 1 & 0 \\ 
1 & 0 & 0 & 0 & 0 \\ 
0 & 1 & 0 & 0 & 0 \\ 
0 & 0 & 0 & 0 & 1%
\end{array}%
\right) ,y^{\prime 4}=\left( 
\begin{array}{ccccc}
0 & 0 & 0 & 1 & 0 \\ 
0 & 0 & 1 & 0 & 0 \\ 
1 & 0 & 0 & 0 & 0 \\ 
0 & 1 & 0 & 0 & 0 \\ 
0 & 0 & 0 & 0 & 1%
\end{array}%
\right) .
\end{align*}%
Then,%
\begin{eqnarray*}
\epsilon \left( \sigma \right) \left( \bar{y}\right) &=&\frac{1}{12}\text{
where }\bar{y}\in Y_{D}\left( y\right) =\left\{ y^{1},\cdots ,y^{4}\right\} ,
\\
\epsilon \left( \sigma \right) \left( \bar{y}^{\prime }\right) &=&\frac{1}{6}%
\text{ where }\bar{y}^{\prime }\in Y_{D}\left( y^{\prime }\right) =\left\{
y^{\prime 1},\cdots ,y^{\prime 4}\right\} .
\end{eqnarray*}%
Under $\epsilon \left( \sigma \right) $, each of agents $a_{1}$ and $a_{2}$
receives one copy of object type $o_{1},o_{2},o_{3}$, and $o_{4}$ with
probabilities $1/6$, $1/6,$ $1/3$, and $1/3$, respectively. Similarly, for
each of $a_{3}$ and $a_{4}$, the individual assignment of $a$ is $%
o_{1},o_{2},o_{3}$, and $o_{4}$ with probabilities $1/3$, $1/3$, $1/6$, and $%
1/6$, respectively.

Thus, $\epsilon \left( \sigma \right) $ satisfies ETE. \newline

We have the following result.

\begin{lemma}
For all $a\in A_{n}$ and all $x\in X$, 
\begin{equation}
\Pr \left( x;\mathbf{x}\left( \epsilon \left( \sigma \right) \right)
_{a}\right) =\frac{1}{\left\vert A_{n}\right\vert }\sum\limits_{b\in
A_{n}}\Pr \left( x;\mathbf{x}\left( \sigma \right) _{b}\right) .  \label{c}
\end{equation}
\end{lemma}

\textbf{Proof. }Fix an arbitrary probabilistic assignment $\sigma $ and an
arbitrary assignment for one agent$\ x\in X$. Moreover, we consider $A_{n}$.

Let $\bar{Y}$ be the set of pure assignments satisfying for any $y\in \bar{Y}
$, $y_{a}=x$ for some $a\in A_{n}$. First, if $\bar{Y}=\emptyset $, then $%
\Pr \left( x;\mathbf{x}\left( \sigma \right) _{a}\right) =0$ and $\Pr \left(
x;\mathbf{x}\left( \epsilon \left( \sigma \right) \right) _{a}\right) =0$
for all $a\in A_{n}$. Therefore, we have (\ref{c}) in the case where $\bar{Y}%
=\emptyset $.

Second, suppose $\bar{Y}\neq \emptyset $ and let $y\in \bar{Y}$. Let $\bar{n}%
:\bar{Y}\rightarrow \left\{ 1,\cdots ,\left\vert A_{n}\right\vert \right\} $
where $\bar{n}\left( y\right) $ represents the number of agents in $A_{n}$
who obtains $x$ at $y\in \bar{Y}$. We consider $Y_{D}\left( y\right) $.
Then, for each agent in $A_{n}$ denoted by $b$, there are $L_{-n}\times \bar{%
n}\left( y\right) \times \left( \left\vert A_{n}\right\vert -1\right) !$
pure assignments in $Y_{D}\left( y\right) $ where $b$ obtains $x$. This
follows because one chooses which of the $\bar{n}\left( y\right) $ agents
receiving $x$ is assigned to $b$, permutes the remaining $\left\vert
A_{n}\right\vert -1$ agents in the group, and combines this with all
permutations of agents in the other groups.

Therefore, for each $a\in A_{n}$, if $\bar{Y}\neq \emptyset $, then%
\begin{eqnarray*}
\Pr \left( x;\mathbf{x}\left( \epsilon \left( \sigma \right) \right)
_{a}\right) &=&\sum\limits_{y\in \bar{Y}}\sigma \left( y\right) \frac{%
L_{-n}\times \bar{n}\left( y\right) \times \left( \left\vert
A_{n}\right\vert -1\right) !}{L} \\
&=&\sum\limits_{y\in \bar{Y}}\sigma \left( y\right) \frac{\bar{n}\left(
y\right) }{\left\vert A_{n}\right\vert }=\frac{1}{\left\vert
A_{n}\right\vert }\sum\limits_{b\in A_{n}}\Pr \left( x;\mathbf{x}\left(
\sigma \right) _{b}\right) .
\end{eqnarray*}%
\textbf{Q.E.D.}\newline

By Lemma 1, we immediately have the following result.

\begin{proposition}
For any $\sigma \in \Sigma $, $\epsilon \left( \sigma \right) $ satisfies
ETE.
\end{proposition}

Since $L$ can be extremely large, explicitly constructing the ETE
reassignment may at first appear computationally demanding. Nevertheless,
Lemma 1 implies that the ETE reassignment of any $\sigma \in \Sigma $ can be
implemented by the following simple procedure. First, draw a pure assignment 
$y$ according to $\sigma $. Second, for each group of equals, pool the
individual assignments received by its members under $y$ and randomly
redistribute them among the members, with each permutation being equally
likely. Thus, provided that an outcome can be sampled from $\sigma $ in
polynomial time, an outcome can also be sampled from $\epsilon \left( \sigma
\right) $ in polynomial time without explicitly enumerating the support of
the resulting distribution.

Next, we consider the ETE reassignment of an EE assignment.

\begin{proposition}
The ETE reassignment of an EE assignment is EE.
\end{proposition}

\textbf{Proof. }Let $y\in \mathrm{Supp}(\sigma )$ be an efficient pure
assignment. Let $y^{l}$ be an arbitrary assignment derived from $y$, where $%
\pi ^{l}$ is the permutation used to construct $y^{l}$ from $y$. We show
that $y^{l}$ is also efficient. Suppose not; that is, $y^{l}$ is Pareto
dominated by a feasible assignment $y^{\prime }$. Let $y^{\prime \prime }$
be such that $y_{ao}^{\prime \prime }=y_{\pi ^{l}(a)o}^{\prime }$ for all $%
a\in A$. Then, by Assumption 2, $y^{\prime \prime }$ is feasible and Pareto
dominates $y$. However, these facts contradict that $y$ is efficient. 
\textbf{Q.E.D.}\newline

By Propositions 1 and 2, the ETE reassignment of an EE assignment satisfies
both EE and ETE. Once an EE assignment is available, obtaining an assignment
that satisfies both EE and ETE is straightforward.

Next, we consider the case where the affirmative action policy is adopted.
For example, suppose that an affirmative action policy is implemented such
that each agent in $A_{n}$ has a characteristic that justifies preferential
treatment under the affirmative action policy, but each agent in $A_{m}$
does not. Note that we allow agents in $A_{n}$ and those in $A_{m}$ to have
identical preferences.

\begin{remark}
Let $\sigma $ be such that for all $a\in A_{n}$ and all $b\in A_{m}$, $%
\mathbf{x}\left( \sigma \right) _{b}$ is first-order stochastically
dominated by $\mathbf{x}\left( \sigma \right) _{a}$ for $a$, and for some $%
a^{\prime }\in A_{n}$ and some $b^{\prime }\in A_{m}$, $\mathbf{x}\left(
\sigma \right) _{b^{\prime }}$ is strictly first-order stochastically
dominated for $a$ by $\mathbf{x}\left( \sigma \right) _{a^{\prime }}$. Then,
for all $a\in A_{n}$ and $b\in A_{m}$, $\mathbf{x}\left( \epsilon \left(
\sigma \right) \right) _{b}$ is strictly first-order stochastically
dominated by $\mathbf{x}\left( \epsilon \left( \sigma \right) \right) _{a}$
for $a$.
\end{remark}

\textbf{Proof. }Fix an agent $a$ and\textbf{\ }let $\left\{
x^{1},x^{2},\cdots ,x^{\left\vert X\right\vert }\right\} =X$ be such that $%
x^{i}\succ _{a}x^{i+1}$ for all $i=1,2,\cdots ,\left\vert X\right\vert -1$.
Then, for all $a\in A_{n}$, all $b\in A_{m}$, all $i=1,2,\cdots ,\left\vert
X\right\vert $, 
\begin{equation*}
\bar{F}_{a}\left( x^{i},\mathbf{x}\left( \sigma \right) _{a}\right) \geq 
\bar{F}_{a}\left( x^{i},\mathbf{x}\left( \sigma \right) _{b}\right) ,
\end{equation*}%
and for some $a^{\prime }\in A_{n}$, some $b^{\prime }\in A_{m},$ and some $%
j=1,2,\cdots ,\left\vert X\right\vert $,%
\begin{equation*}
\bar{F}_{a}\left( x^{j},\mathbf{x}\left( \sigma \right) _{a^{\prime
}}\right) >\bar{F}_{a}\left( x^{j},\mathbf{x}\left( \sigma \right)
_{b^{\prime }}\right) .
\end{equation*}%
By (\ref{c}), 
\begin{eqnarray*}
\Pr \left( x;\mathbf{x}\left( \epsilon \left( \sigma \right) \right)
_{a}\right) &=&\frac{1}{\left\vert A_{n}\right\vert }\sum\limits_{\hat{a}\in
A_{n}}\Pr \left( x;\mathbf{x}\left( \sigma \right) _{\hat{a}}\right) , \\
\Pr \left( x;\mathbf{x}\left( \epsilon \left( \sigma \right) \right)
_{b}\right) &=&\frac{1}{\left\vert A_{m}\right\vert }\sum\limits_{\hat{b}\in
A_{m}}\Pr \left( x;\mathbf{x}\left( \sigma \right) _{\hat{b}}\right) .
\end{eqnarray*}%
Therefore, for all $a\in A_{n}$, all $b\in A_{m}$, all $i=1,2,\cdots
,\left\vert X\right\vert $,%
\begin{equation*}
\bar{F}_{a}\left( x^{i},\mathbf{x}\left( \epsilon \left( \sigma \right)
\right) _{a}\right) \geq \bar{F}_{a}\left( x^{i},\mathbf{x}\left( \epsilon
\left( \sigma \right) \right) _{b}\right) .
\end{equation*}%
Moreover, since the strict inequality holds for some pair $\left( a^{\prime
},b^{\prime }\right) ,$ averaging preserves strictness, so%
\begin{equation*}
\bar{F}_{a}\left( x^{j},\mathbf{x}\left( \epsilon \left( \sigma \right)
\right) _{a^{\prime }}\right) >\bar{F}_{a}\left( x^{j},\mathbf{x}\left(
\epsilon \left( \sigma \right) \right) _{b^{\prime }}\right) .
\end{equation*}%
Therefore, for all $a\in A_{n}$ and $b\in A_{m}$, $\mathbf{x}\left( \epsilon
\left( \sigma \right) \right) _{b}$ is strictly first-order stochastically
dominated by $\mathbf{x}\left( \epsilon \left( \sigma \right) \right) _{a}$
for $a$. \textbf{Q.E.D.}\newline

We consider an assignment $\sigma $ such that agents in $A_{n}$ receive
preferential treatment compared to those in $A_{m}$. Then, this property is
preserved in the ETE reassignment of $\sigma $. Thus, the ETE reassignment
allows us to obtain an assignment that satisfies ETE while maintaining
preferential treatment for agents possessing specific characteristics even
if they have identical preferences. Note that this does not happen when we
simply treat agents with identical preferences as equals, as defined in
previous studies.

\section{Ordinal Efficiency and Rank-minimizing}

In this section, we focus on two efficiency notions that are stronger than
EE. First, an assignment $\sigma $ is said to be \textbf{ordinally dominated}
by $\sigma ^{\prime }$ if $\mathbf{x}\left( \sigma \right) _{a}$ is
first-order stochastically dominated by $\mathbf{x}\left( \sigma ^{\prime
}\right) _{a}$ for all $a\in A$ and the former strictly first-order
stochastically dominated by the latter for some $a\in A$. An assignment is
said to be \textbf{ordinally efficient} (\textbf{OE}) if it is not ordinally
dominated by any other assignment. Note that a pure assignment is OE if and
only if it is efficient.

First, we have the following result.

\begin{lemma}
Let $\sigma $ be an OE assignment and let $y\in \mathrm{Supp}\left( \sigma
\right) $. Then, $\sigma ^{y}$ is also OE.
\end{lemma}

\textbf{Proof. }Let $\sigma $ be an OE assignment. Suppose not; that is, $%
y\in \mathrm{Supp}\left( \sigma \right) $ such that $\sigma ^{y}$ is not OE.
Then, there is an assignment $\sigma ^{\ast }$ that ordinally dominates $%
\sigma ^{y}$. Since 
\begin{equation*}
\sum\limits_{\hat{y}\in Y}\sigma ^{y}\left( \hat{y}\right) =1\text{ and }%
\sum\limits_{\hat{y}\in Y}\sigma \left( \hat{y}\right) =1,
\end{equation*}%
we have 
\begin{equation*}
\sigma \left( y\right) \times \sum\limits_{\hat{y}\in Y}\sigma ^{\ast
}\left( \hat{y}\right) +\sum_{y^{\prime }\in Y\setminus \left\{ y\right\}
}\sigma \left( y^{\prime }\right) =1.
\end{equation*}%
Define an assignment $\sigma ^{\ast \ast }$ by%
\begin{gather*}
\sigma ^{\ast \ast }\left( y\right) =\sigma \left( y\right) \times \sigma
^{\ast }\left( y\right) \\
\sigma ^{\ast \ast }\left( y^{\prime }\right) =\sigma \left( y\right) \times
\sigma ^{\ast }\left( y^{\prime }\right) +\sigma \left( y^{\prime }\right) ,
\end{gather*}%
for all $y^{\prime }\in Y\setminus \left\{ y\right\} $. Then, $\sigma ^{\ast
\ast }$ is a valid assignment. Since $\sigma ^{\ast }$ ordinally dominates $%
\sigma ^{y}$ and $\sigma \left( y\right) >0$, $\sigma ^{\ast \ast }$
ordinally dominates $\sigma $. However, this contradicts that $\sigma $ is
OE. \textbf{Q.E.D.}\newline

In this study, we represent an assignment as a lottery over pure
assignments. Lemma 2 means that if a lottery over pure assignments is OE,
then the deterministic assignment realizing one of them with probability 1
is also OE. This immediately implies the following fact.

\begin{corollary}
An OE assignment is EE.
\end{corollary}

\subsection{Simple-constraint case}

Contrary to Lemma 2, Bogomolnaia and Moulin (2001) show that, even in the
simple-constraint case, an assignment may fail to be OE although all pure
assignments in its support are OE. Nevertheless, as shown below, while the
ETE reassignment of an OE assignment need not be OE in general, it is always
OE in the simple-constraint case.

\begin{proposition}
In the simple-constraint case, the ETE reassignment of an OE assignment is
also OE.
\end{proposition}

This result follows from Theorem 1 in Okumura (2026b), which builds on
Proposition 5 of Kesten and \"{U}nver (2015). Okumura (2026b) considers a
model in which object types, called schools in his terminology, have
priority orders over agents, called students in his terminology, and shows
that the ETE reassignment of a constrained efficient deterministic
assignment is an ex ante stable lottery that is not ordinally dominated by
any other ex ante stable lottery. The simple-constraint case corresponds to
the special case of his model in which all priority relations are completely
indifferent. In this case, every feasible assignment is ex ante stable, and
constrained efficiency coincides with ordinal efficiency. Hence, the ETE
reassignment of any OE assignment is also OE.

However, in general, the ETE reassignment of an OE assignment may fail to be
OE. We provide the following example.

\subsubsection*{Example 3}

Let $A=\left\{ a_{1},\cdots ,a_{4}\right\} $, $O=\left\{ o_{1},\cdots
,o_{4}\right\} $, and all agents are equals in this example. Here, we assume
that $Y$ satisfies\textbf{\ }single-unit demand. Moreover, suppose 
\begin{equation*}
\succ _{a}:o_{1},o_{2},o_{3},o_{4}
\end{equation*}%
for all $a\in A$.\footnote{%
This means that $o_{1}\succ _{a}o_{2}\succ _{a}o_{3}\succ _{a}o_{4}$ for all 
$a\in A$. In the examples in this study, the preferences of agents are
represented in this simplified manner.} Let%
\begin{eqnarray*}
y &=&\left( 
\begin{array}{cccc}
y_{a_{1}o_{1}} & y_{a_{1}o_{2}} & y_{a_{1}o_{3}} & y_{a_{1}o_{4}} \\ 
y_{a_{2}o_{1}} & y_{a_{2}o_{2}} & y_{a_{2}o_{3}} & y_{a_{2}o_{4}} \\ 
y_{a_{3}o_{1}} & y_{a_{3}o_{2}} & y_{a_{3}o_{3}} & y_{a_{3}o_{4}} \\ 
y_{a_{4}o_{1}} & y_{a_{4}o_{2}} & y_{a_{4}o_{3}} & y_{a_{4}o_{4}}%
\end{array}%
\right) =\left( 
\begin{array}{cccc}
1 & 0 & 0 & 0 \\ 
0 & 1 & 0 & 0 \\ 
0 & 0 & 1 & 0 \\ 
0 & 0 & 0 & 1%
\end{array}%
\right) , \\
y^{\prime } &=&\left( 
\begin{array}{cccc}
1 & 0 & 0 & 0 \\ 
1 & 0 & 0 & 0 \\ 
0 & 0 & 0 & 1 \\ 
0 & 0 & 0 & 1%
\end{array}%
\right) ,y^{\prime \prime }=\left( 
\begin{array}{cccc}
0 & 1 & 0 & 0 \\ 
0 & 1 & 0 & 0 \\ 
0 & 1 & 0 & 0 \\ 
0 & 1 & 0 & 0%
\end{array}%
\right) .
\end{eqnarray*}%
We assume that a pure assignment is feasible if and only if it is derived
from $y,$ $y^{\prime }$ or $y^{\prime \prime }$. Then, $\sigma ^{y}$ is an
OE assignment.

Then, 
\begin{equation*}
\Pr \left( o_{i};\mathbf{x}\left( \epsilon \left( \sigma ^{y}\right) \right)
_{a}\right) =\frac{1}{4}
\end{equation*}%
for all $i=1,\cdots ,4$ and all $a\in A$.

Let%
\begin{equation*}
y^{1}=\left( 
\begin{array}{cccc}
0 & 0 & 0 & 1 \\ 
0 & 0 & 1 & 0 \\ 
1 & 0 & 0 & 0 \\ 
0 & 1 & 0 & 0%
\end{array}%
\right) ,y^{2}=\left( 
\begin{array}{cccc}
0 & 1 & 0 & 0 \\ 
0 & 0 & 0 & 1 \\ 
0 & 0 & 1 & 0 \\ 
1 & 0 & 0 & 0%
\end{array}%
\right) ,
\end{equation*}%
which are derived from $y$ and thus feasible. Define $\sigma ^{\prime }$ by 
\begin{equation*}
\sigma ^{\prime }\left( y^{\prime }\right) =\sigma ^{\prime }\left(
y^{\prime \prime }\right) =\sigma ^{\prime }\left( y^{1}\right) =\sigma
^{\prime }\left( y^{2}\right) =\frac{1}{4}.
\end{equation*}%
Then, for all $a\in \left\{ a_{1},a_{4}\right\} $,%
\begin{eqnarray*}
\Pr \left( o_{1};\mathbf{x}\left( \sigma ^{\prime }\right) _{a}\right)
&=&\Pr \left( o_{4};\mathbf{x}\left( \sigma ^{\prime }\right) _{a}\right) =%
\frac{1}{4}, \\
\Pr \left( o_{2};\mathbf{x}\left( \sigma ^{\prime }\right) _{a}\right) &=&%
\frac{1}{2},\Pr \left( o_{3};\mathbf{x}\left( \sigma ^{\prime }\right)
_{a}\right) =0,
\end{eqnarray*}%
and for all $a^{\prime }\in \left\{ a_{2},a_{3}\right\} $ and all $%
i=1,\cdots ,4$, 
\begin{equation*}
\Pr \left( o_{i};\mathbf{x}\left( \sigma ^{\prime }\right) _{a^{\prime
}}\right) =\frac{1}{4}.
\end{equation*}

Then, since $\mathbf{x}\left( \epsilon \left( \sigma ^{y}\right) \right)
_{a} $ is first-order stochastically dominated by $\mathbf{x}\left( \sigma
^{\prime }\right) _{a}$ for all $a\in A$, with strict inequality for $a_{1}$
and $a_{4}$, $\epsilon \left( \sigma ^{y}\right) $ is not OE, even though $%
\sigma ^{y}$ is an OE assignment.\newline

Thus, there may exist some OE assignment whose ETE reassignment is not OE.
In the next subsection, we show that there exists an OE assignment whose ETE
reassignment is also OE.

\subsection{General existence result}

Here, we consider whether there exists an OE assignment whose ETE
reassignment is also OE. To address this question, we introduce a stronger
notion of efficiency than OE.

For each $a\in A$, let 
\begin{equation*}
r\left( x;a\right) =\left\vert \left\{ x^{\prime }\in X\text{ }\left\vert 
\text{ }x^{\prime }\succ _{a}x\right. \right\} \right\vert +1,
\end{equation*}%
which represents the rank position of $x$ in the preference order of $a$.
Note that $\bar{F}_{a}\left( x,\mathbf{x}\right) $ represents the
probability that the rank of the object type assigned to $a$ is $r\left(
x;a\right) $ or better. Let $R:\Sigma \rightarrow \mathbb{R}$ be defined by%
\begin{equation*}
R\left( \sigma \right) =\sum\limits_{y\in Y}\sigma \left( y\right)
\sum\limits_{a\in A}r\left( y_{a};a\right) ,
\end{equation*}%
which is the expected value of the sum of rank positions of the assignment
in the preference orders of all agents. We say that $\sigma \in \Sigma $ is 
\textbf{rank-minimizing efficient (RE)} if $R\left( \sigma \right) \leq
R\left( \sigma ^{\prime }\right) $ for all $\sigma ^{\prime }\in \Sigma $.

Featherstone (2020) shows that there exist RE assignments and each of them
is OE in the simple-constraint case. We generalize the result.

\begin{lemma}
First, there exists some RE assignment. Second, any RE assignment is OE.
\end{lemma}

\textbf{Proof.} Since $\Sigma $ is compact and $R$ is continuous on $\Sigma $%
, the existence of some RE assignments is trivial. We show that each of them
is OE. We use the following result.

\begin{claim}
If $\sigma \in \Sigma $ is ordinally\textbf{\ }dominated by $\sigma ^{\prime
}\in \Sigma $, then $R\left( \sigma \right) >R\left( \sigma ^{\prime
}\right) $.
\end{claim}

\textbf{Proof of Claim 1. }Since $\mathbf{x}\left( \sigma \right) _{a}$ is
first-order stochastically dominated by $\mathbf{x}\left( \sigma ^{\prime
}\right) _{a}$ for all $a$, 
\begin{equation*}
\sum\limits_{y\in Y}\sigma \left( y\right) r\left( y_{a};a\right) \geq
\sum\limits_{y\in Y}\sigma ^{\prime }\left( y\right) r\left( y_{a};a\right) .
\end{equation*}%
Moreover, since the dominance is strict for $a^{\prime }$, 
\begin{equation*}
\sum\limits_{y\in Y}\sigma \left( y\right) r\left( y_{a^{\prime }};a^{\prime
}\right) >\sum\limits_{y\in Y}\sigma ^{\prime }\left( y\right) r\left(
y_{a^{\prime }};a^{\prime }\right) .
\end{equation*}%
Thus,%
\begin{equation*}
R\left( \sigma \right) =\sum\limits_{a\in A}\sum\limits_{y\in Y}\sigma
\left( y\right) r\left( y_{a};a\right) >\sum\limits_{a\in
A}\sum\limits_{y\in Y}\sigma ^{\prime }\left( y\right) r\left(
y_{a};a\right) =R\left( \sigma ^{\prime }\right) \text{.}
\end{equation*}%
\textbf{Q.E.D.}

By Claim 1, $R\left( \sigma \right) \leq R\left( \sigma ^{\prime }\right) $
for any $\sigma ^{\prime }\in \Sigma $ implies that $\sigma $ is not
ordinally\textbf{\ }dominated by any $\sigma ^{\prime }\in \Sigma $. \textbf{%
Q.E.D.}\newline

Thus, there always exists an RE assignment that must be OE. As a further
advantage of this efficiency notion, we have the following result.

\begin{theorem}
If $\sigma $ is RE, then $\epsilon \left( \sigma \right) $ is also RE. Thus,
there must exist an assignment that satisfies both ETE and RE, and thus OE.%
\footnote{%
In Example 3, $\sigma ^{y^{\prime \prime }}$ is an RE assignment and
satisfies ETE}
\end{theorem}

\textbf{Proof.} We show that the ETE reassignment of an RE assignment must
be RE. To show this result, we prove the following result.

\begin{lemma}
An assignment $\sigma $ is RE if and only if for all pure assignment $y\in 
\mathrm{Supp}\left( \sigma \right) ,$ $\sigma ^{y}$ is RE.
\end{lemma}

\textbf{Proof.} We arbitrarily choose $y\in \mathrm{Supp}\left( \sigma
\right) $, where $\sigma $ is an RE assignment. First, we show that the pure
assignment $\sigma ^{y}$ is also an RE assignment.

Suppose not; that is,

\begin{equation*}
R\left( \sigma \right) =\sum\limits_{y\in Y}\sigma \left( y\right)
\sum\limits_{a\in A}r\left( y_{a};a\right) <\sum\limits_{a\in A}r\left(
y_{a};a\right) \text{.}
\end{equation*}%
Then, since $\sigma \left( y\right) >0$, there must exist $y^{\prime }\in 
\mathrm{Supp}\left( \sigma \right) $ such that 
\begin{equation*}
R\left( \sigma ^{y^{\prime }}\right) =\sum\limits_{a\in A}r\left(
y_{a}^{\prime };a\right) <\sum\limits_{y\in Y}\sigma \left( y\right)
\sum\limits_{a\in A}r\left( y_{a};a\right) ,
\end{equation*}%
which contradicts that $\sigma $ is an RE assignment. Therefore, 
\begin{equation}
R\left( \sigma \right) =R\left( \sigma ^{y}\right) =\sum\limits_{a\in
A}r\left( y_{a};a\right) ;  \label{a}
\end{equation}
that is, $\sigma ^{y}$ is also an RE assignment.

Conversely, suppose that $\mathrm{Supp}\left( \sigma \right) =\left\{
y^{1},y^{2},\cdots ,y^{n}\right\} $ and that $\sigma ^{y^{i}}$ is RE for all 
$i=1,\cdots ,n$. Then, we can let 
\begin{equation*}
R^{\ast }=R\left( \sigma ^{y^{1}}\right) =\cdots =R\left( \sigma
^{y^{n}}\right) \text{.}
\end{equation*}%
for some constant $R^{\ast }$. Since 
\begin{equation*}
R\left( \sigma ^{y^{i}}\right) =\sum\limits_{a\in A}r\left(
y_{a}^{i};a\right)
\end{equation*}%
for all $i=1,\cdots ,n,$ 
\begin{equation*}
R\left( \sigma \right) =\sum\limits_{y\in Y}\sigma \left( y\right) R^{\ast
}=R^{\ast }=R\left( \sigma ^{y^{i}}\right) \text{,}
\end{equation*}%
and thus $\sigma $ is RE. \textbf{Q.E.D.}\newline

We now prove the first claim of the theorem. Let $\sigma $ be an RE
assignment and$\ y\in \mathrm{Supp}\left( \sigma \right) $. By Lemma 4, 
\begin{equation*}
R\left( \sigma \right) =\sum\limits_{a\in A}r\left( y_{a};a\right) .
\end{equation*}%
By construction, for any $y^{\prime }\in Y_{D}\left( y\right) $, 
\begin{equation*}
\sum\limits_{a\in A}r\left( y_{a}^{\prime };a\right) =\sum\limits_{a\in
A}r\left( y_{a};a\right) .
\end{equation*}%
Hence, every pure assignment in the support of $\epsilon \left( \sigma
\right) $ attains the same total rank. By Lemma 4, $\epsilon \left( \sigma
\right) $ is also RE. Thus, we have the first sentence of Theorem 1.

Second, we show the second sentence of Theorem 1. By Lemma 3, there must
exist an RE assignment $\sigma $. By Proposition 1 and the first sentence of
Theorem 1, $\epsilon \left( \sigma \right) $ satisfies ETE and RE. Moreover,
by Lemma 3, $\epsilon \left( \sigma \right) $ also satisfies OE. \textbf{%
Q.E.D.}\newline

As stated above, $Y$ can be taken as the set of matchings that are stable in
the ex-post sense with the priority orders of object types. In this case,
Theorem 1 establishes the existence of an ETE assignment whose total rank is
no greater than that of any other stable matching. However, under Assumption
2, this implicitly requires each school's priority order to treat equals as
tied.\footnote{%
Kesten and \"{U}nver (2015), Han (2024) and Okumura (2026b) explicitly
incorporate school priorities into their definitions of equals, regarding
two students as equals if and only if they have identical preferences and
are tied in priority at every school.}

In what follows, we examine whether such an assignment can be derived in a
computationally efficient manner. First, in the simple-constraint case, an
RE assignment can be computed efficiently (see, for example, Korte and Vygen
2005, Ch. 11). Since an ETE reassignment for a given assignment can be
performed in polynomial time, we can obtain an RE assignment that satisfies
ETE in a computationally efficient way in this specific case.

As noted above, the literature has introduced the uniform RE mechanism,
which implements all RE assignments with equal probability. However, as
pointed out by Troyan (2024), finding all RE assignments is computationally
infeasible even in the simple-constraint case, which raises a concern from
the perspective of computational complexity. In contrast, our method derives
an assignment satisfying both ETE and RE in a computationally efficient
manner in this setting.

However, in more general settings, no computationally efficient method is
known for deriving an assignment that satisfies both ETE and RE---or even
the weaker requirement of OE. Although, as is clear from Lemma 1, the ETE
reassignment of any given assignment can be implemented efficiently, no
computationally efficient method is known for finding an RE assignment in
the general case. Moreover, even after weakening RE to OE, it remains
unclear how to efficiently construct an assignment satisfying both ETE and
OE. Therefore, in the next subsection, we propose such a construction for a
broad class of environments.

\subsection{General upper bounds}

Here, we consider the case in which $Y$ satisfies the following conditions.
The set of feasible pure assignments $Y$ satisfies the \textbf{general upper
bounds constraint} if for any $y\in Y,$ any $y^{\prime }\in \mathbb{Z}%
_{+}^{\left\vert A\right\vert \times \left\vert O\right\vert }$ such that $%
0\leq y_{ao}^{\prime }\leq y_{ao}$ for all $\left( a,o\right) \in \left(
A\times O\right) $ also belongs to $Y$.

We introduce a specific version of this constraint. For each $o\in O$, let $%
Z_{o}\subseteq \mathbb{Z}_{+}^{\left\vert A\right\vert }$ denote the set of
feasible assignments for $o$. We say that $Z_{o}$ satisfies the \textbf{%
general upper bounds constraint for }$o$\textbf{\ }if, for any $z\in Z_{o},$
any $z^{\prime }\in \mathbb{Z}_{+}^{\left\vert A\right\vert }$ satisfying $%
z_{a}^{\prime }\leq z_{a}$ also belongs to $Z_{o}$. We say that the set of
feasible pure assignments $Y$ satisfies the \textbf{general upper bounds
constraint for each object type}, if for every $o\in O$, the set $%
Z_{o}\subseteq \mathbb{Z}_{+}^{\left\vert A\right\vert }$ satisfies the
general upper bounds constraint for\textbf{\ }$o$, and any $y\in \mathbb{Z}%
_{+}^{\left\vert A\right\vert \times \left\vert O\right\vert }$ such that $%
y_{o}\in Z_{o}$ for all $o\in O$ belongs to $Y$.

As shown in the example below, the former constraint structure is more
general than the latter; that is, if $Y$ satisfies the general upper bounds
constraint for each object type, then it also satisfies the general upper
bounds constraint.

Okumura (2019) and Kamada and Kojima (2024) consider the set of (pure)
assignments satisfying the general upper bounds constraint for each object
type in the unit-demand case. Imamura and Kawase (2025) study the set of
assignments that satisfy the general upper bounds constraint, though their
model is also restricted to the unit-demand case.

We consider the difference of these constraints by introducing an example.

\subsubsection*{Example 4}

Let $A=\left\{ a_{1},a_{2}\right\} $ and $O=\left\{ o_{1},o_{2}\right\} $.
Let 
\begin{eqnarray*}
y &=&\left( 
\begin{array}{cc}
y_{a_{1}o_{1}} & y_{a_{1}o_{2}} \\ 
y_{a_{2}o_{1}} & y_{a_{2}o_{2}}%
\end{array}%
\right) =\left( 
\begin{array}{cc}
1 & 0 \\ 
0 & 0%
\end{array}%
\right) , \\
y^{\prime } &=&\left( 
\begin{array}{cc}
0 & 0 \\ 
0 & 1%
\end{array}%
\right) ,y^{\prime \prime }=\left( 
\begin{array}{cc}
1 & 0 \\ 
0 & 1%
\end{array}%
\right) .
\end{eqnarray*}%
Suppose $y,y^{\prime }\in Y$. If $Y$ satisfies the general upper bounds
constraint for each object type, then $y^{\prime \prime }\in Y$. However, if 
$Y$ satisfies only the general upper bounds constraint, then it is possible
that $y^{\prime \prime }\notin Y$.\newline

To illustrate the importance of the general upper bounds constraint, we
consider the following examples.

First, as discussed by Kamada and Kojima (2015), regional maximum quotas
have been introduced in the Japanese medical residency matching market to
mitigate the overconcentration in specific regions. Such regional
limitations are incorporated into our model through general upper bound
constraints. To see this point, we revisit Example 3. Suppose that $o_{1}$
and $o_{2}$ are in the same region, and there is the regional maximum quota;
that is, $y\in Y$ if and only if%
\begin{equation*}
\sum\limits_{a\in A}\left( y_{ao_{1}}+y_{ao_{2}}\right) \leq 1.
\end{equation*}%
Then, $y,y^{\prime }\in Y$ but $y^{\prime \prime }\notin Y$.

Similarly, in order to achieve the same objective of ensuring a certain
level of allocation to specific regions, regional minimum quotas can also be
introduced. Specifically, suppose that only $o_{1}$ and $o_{2}$ belong to
the same region, and that this region has a minimum quota of $n\in \left(
0,\left\vert A\right\vert \right) $. Thus, an assignment $y$ is feasible if
and only if the following condition holds: 
\begin{equation}
\sum\limits_{a\in A}\left( y_{ao_{1}}+y_{ao_{2}}\right) \geq n.  \label{f}
\end{equation}%
At first glance, this may appear to be incompatible with general upper bound
constraints. However, as pointed out by Balbuzanov (2022), such minimum
quota constraints can in fact be reformulated and treated as general upper
bound constraints.

Specifically, we assume single-unit demand and that there exists a null
object type $o_{0}\in O$, corresponding to the outside option of each agent.
Under this formulation, constraint (\ref{f}) is equivalent to 
\begin{equation*}
\sum\limits_{a\in A,o\in O\setminus \left\{ o_{1},o_{2}\right\} }y_{ao}\leq
\left\vert A\right\vert -n.
\end{equation*}%
Therefore, assignment problems with regional minimum quotas are also
encompassed by the case where $Y$ satisfies general upper bound constraints.%
\footnote{%
Balbuzanov (2022) shows that any feasibility constraint can be reformulated
as an upper-bound constraint; however, this transformation may not be
computable in polynomial time for certain classes of constraints. Since this
section of this paper focuses on computationally efficient methods, the
approach proposed by Balbuzanov (2022) cannot be generally employed, except
in trivial cases.} Note that this reformulation relies on the presence of
the null object type, which absorbs the remaining assignments.

Second, we consider the case where each school may have multiple admission
slots with different amounts of scholarship. Suppose that there is one
school with a total capacity of 200 students. The school offers three types
of admission slots: one with a scholarship of 4,000 USD, denoted by $o_{1}$;
another with a scholarship of 2,000 USD, denoted by $o_{2}$; and one without
any scholarship, denoted by $o_{3}$. We assume that the school has a total
scholarship financial budget of 100,000 USD. Then, $y\in Y$ if and only if 
\begin{eqnarray*}
\sum_{a\in A}\sum_{o\in \left\{ o_{1},o_{2},o_{3}\right\} }y_{ao} &\leq &200,
\\
\sum_{a\in A}y_{ao_{1}}\times 4000+\sum_{a\in A}y_{ao_{2}}\times 2000 &\leq
&100,000.
\end{eqnarray*}%
This $Y$ satisfies the general upper bounds constraint, but does not satisfy
the general upper bounds constraint for each object type.

Third, we consider a controlled school choice problem. There is also one
school with a total capacity of 200 students. Let $S_{1}$, $S_{2}$, and $%
S_{3}$ be the sets of students (agents in our terminology) from different
categories, respectively. Note that the intersection of any two of these
three sets may be non-empty. This implies that, as also considered by Kurata
et al. (2017), Ayg\"{u}n and Bo (2021), and S\"{o}nmez and Yenmez (2022), we
allow for cases in which an agent possesses multiple characteristics that
justify preferential treatment under affirmative action. As stated by Ayg%
\"{u}n and Turhan (2017, 2020), students from these categories may have
preferences not only over schools but also over the admission categories
through which they are accepted, as exemplified by the case of engineering
school admissions in India.

Based on the case of engineering school admissions in India, there exist
four different slots as below. Let $o_{1}$, $o_{2}$, and $o_{3}$ be reserved
slots for students in $S_{1}$, $S_{2}$, and $S_{3},$ respectively. Moreover,
let $o_{4}$ be the open slot that can be assigned to any students. We
consider the hard bound constraints; that is, if there are not enough
applications for $o_{i}$, some seats in $o_{i}$ will remain empty for $%
i=1,2,3$.\footnote{%
In the actual case of engineering school admissions in India, the capacities
of some reserved slots are treated as hard bounds---if there are not enough
applicants, those slots remain vacant. However, some other types of reserved
slots may be converted into general-category slots if left vacant. For
further details, see Ayg\"{u}n and Turhan (2017).} Then, for example, let%
\begin{eqnarray*}
\sum_{a\in S_{1}}y_{ao_{1}}+\sum_{a\in A\setminus S_{1}}y_{ao_{1}}\times
\left( \infty \right) &\leq &30, \\
\sum_{a\in S_{2}}y_{ao_{2}}+\sum_{a\in A\setminus S_{2}}y_{ao_{2}}\times
\left( \infty \right) &\leq &15, \\
\sum_{a\in S_{3}}y_{ao_{3}}+\sum_{a\in A\setminus S_{3}}y_{ao_{3}}\times
\left( \infty \right) &\leq &54, \\
\sum_{a\in A}y_{ao_{4}} &\leq &101,
\end{eqnarray*}%
where $\infty $ represents a sufficiently large number. This implies that $%
30,$ $15$, and $54$ seats are reserved for $S_{1},$ $S_{2},$ and $S_{3},$
respectively, and moreover, $101$ seats are open to all students. Note that
students from the categories may be able to choose one from multiple slots.

Even if $Y$ satisfies the general upper bounds constraint for each object
type, the ETE reassignment of an OE assignment may not be OE. To illustrate
this fact, we provide the following example.\footnote{%
This example (Example 5) is suggested by Minoru Kitahara, and I am grateful
for his contribution.}

\subsubsection*{Example 5}

We consider the unit-demand case. Let $A=A_{1}\cup A_{2}$ where $%
A_{1}=\left\{ a_{1},a_{2},a_{3}\right\} $ and $A_{2}=\left\{
a_{4},a_{5},a_{6}\right\} $, and $O=\left\{ o_{1}\right\} $. Suppose that $%
y\in Y$ if either (1) $\sum\nolimits_{a\in A}y_{ao_{1}}\leq 2$ or (2) $%
y_{ao_{1}}=0$ for all $a\in A_{1}$ or all $a\in A_{2}$. That is, three
agents can simultaneously obtain one copy of $o_{1}$ if and only if they are
equals. Since there is only one object type, $Y$ satisfies the general upper
bounds constraint for each object type. We assume that every agent prefers
assigning one unit of $o_{1}$ to nothing.

Then, let $y$ be such that $y_{a_{1}o_{1}}=y_{a_{4}o_{1}}=1$ and $%
y_{a_{i}o_{1}}=0$ for all $i=2,3,5,6$. Then, $\epsilon \left( \sigma
^{y}\right) $ satisfies 
\begin{equation*}
\Pr \left( o_{1};\mathbf{x}\left( \epsilon \left( \sigma ^{y}\right) \right)
_{a_{i}}\right) =\frac{1}{3}
\end{equation*}%
for all $i=1,\cdots ,6$. On the other hand, let $\sigma ^{\prime }$ be such
that $\sigma ^{\prime }\left( y^{\prime }\right) =\sigma ^{\prime }\left(
y^{\prime \prime }\right) =1/2$ where%
\begin{eqnarray*}
y_{a_{1}o_{1}}^{\prime } &=&y_{a_{2}o_{1}}^{\prime }=y_{a_{3}o_{1}}^{\prime
}=1,\text{ }y_{a_{4}o_{1}}^{\prime }=y_{a_{5}o_{1}}^{\prime
}=y_{a_{6}o_{1}}^{\prime }=0, \\
y_{a_{1}o_{1}}^{\prime \prime } &=&y_{a_{2}o_{1}}^{\prime \prime
}=y_{a_{3}o_{1}}^{\prime \prime }=0,\text{ }y_{a_{4}o_{1}}^{\prime \prime
}=y_{a_{5}o_{1}}^{\prime \prime }=y_{a_{6}o_{1}}^{\prime \prime }=1.
\end{eqnarray*}%
Then, 
\begin{equation*}
\Pr \left( o_{1};\mathbf{x}\left( \sigma ^{\prime }\right) _{a_{i}}\right) =%
\frac{1}{2},
\end{equation*}%
for all $i=1,\cdots ,6$. Thus, $\epsilon \left( \sigma ^{y}\right) $ is
ordinally dominated by $\sigma ^{\prime }$ even though $\sigma ^{y}$ is an
OE assignment.\newline

Although the ETE reassignment of any OE assignment remains OE in the
simple-constraint case, Example 5 shows that this property fails under
general upper bound constraints; that is, the ETE reassignment of an OE
assignment may fail to be OE. Nevertheless, Theorem 1 guarantees that even
in the complex-constraint case, there exists at least one OE assignment
whose ETE reassignment is also OE. In what follows, we propose a
computationally efficient method to derive such an assignment under general
upper bound constraints.

We introduce the serial dictatorship rules. First, let a priority list $%
\alpha =\left( \alpha _{1},\cdots ,\alpha _{\left\vert A\right\vert }\right) 
$ be a permutation of $A$. The serial dictatorship rule with priority list $%
\alpha $ is as follows:

\begin{description}
\item[Step 0] Let $y^{0}\in Y$ be such that $y_{ao}^{0}=0$ for all $(a,o)\in
A\times O$.

\item[Step $t=1,\cdots ,\left\vert A\right\vert $] Let $y^{t}\in Y$ be such
that $y_{a}^{t}=y_{a}^{t-1}$ for all $a\in \left( A\setminus \left\{ \alpha
_{t}\right\} \right) $, and 
\begin{equation*}
y_{\alpha _{t}}^{t}\in \arg \max_{\succsim _{\alpha _{t}}}\left\{ x\in X%
\text{ }\left\vert \text{ }\exists y\in Y\text{ s.t. }y_{\alpha _{t}}=x,%
\text{ }y_{a}=y_{a}^{t-1}\text{ }\forall a\in A\setminus \left\{ \alpha
_{t}\right\} \right. \right\} .
\end{equation*}
\end{description}

The serial dictatorship rule can be implemented in polynomial time.\footnote{%
Computational efficiency is evaluated under the assumption that individual
assignments and agents' preference orders over them are explicitly
represented and that feasibility can be checked in polynomial time. Under
this representation, serial dictatorship is polynomial-time computable.}
Moreover, we have the following result.

\begin{proposition}
Let $y$ be the result of the serial dictatorship rule with an arbitrary
priority list. If $Y$ satisfies the general upper bounds constraint, then $%
\sigma ^{y}$ is OE.
\end{proposition}

\textbf{Proof. }Let $\alpha $ be an arbitrary priority list. Suppose not;
that is, $\sigma ^{y}$ is ordinally dominated by some assignment $\sigma
^{\prime }$. Then, there is $t\in \left\{ 1,\cdots ,\left\vert A\right\vert
\right\} $ such that 
\begin{equation*}
\Pr \left( y_{\alpha _{t^{\prime }}};\mathbf{x}\left( \sigma ^{y}\right)
_{\alpha _{t^{\prime }}}\right) =1=\Pr \left( y_{\alpha _{t^{\prime }}};%
\mathbf{x}\left( \sigma ^{\prime }\right) _{\alpha _{t^{\prime }}}\right)
\end{equation*}%
for all $t^{\prime }=1,2,\cdots ,t-1,$ and 
\begin{equation*}
\Pr \left( y_{\alpha _{t}};\mathbf{x}\left( \sigma ^{y}\right) _{\alpha
_{t}}\right) =1>\Pr \left( y_{\alpha _{t}};\mathbf{x}\left( \sigma ^{\prime
}\right) _{\alpha _{t}}\right) .
\end{equation*}%
That is, there is $y^{\prime }\in \mathrm{Supp}\left( \sigma ^{\prime
}\right) $ such that $y_{\alpha _{t^{\prime }}}^{\prime }=y_{\alpha
_{t^{\prime }}}$ for all $t^{\prime }=1,2,\cdots ,t-1,$ and $y_{\alpha
_{t}}^{\prime }\succ _{\alpha _{t}}y_{\alpha _{t}}$. Since $Y$ satisfies the
general upper bounds constraint and $y^{\prime }$ is feasible, $y^{\prime
\prime }$ such that $y_{\alpha _{t^{\prime }}}^{\prime \prime }=y_{\alpha
_{t^{\prime }}},$ $y_{\alpha _{t}}^{\prime \prime }=y_{\alpha _{t}}^{\prime
} $ and $y_{\alpha _{t^{\prime \prime }}o}^{\prime \prime }=0$ for all $%
t^{\prime \prime }=t+1,\cdots ,\left\vert A\right\vert $ and $o\in O$ is
also feasible. This implies that, at step $t$, agent $\alpha _{t}$ could
have chosen $y_{\alpha _{t}}^{\prime }$, which they strictly prefer to $%
y_{\alpha _{t}}$, while maintaining feasibility. This contradicts the
construction of the serial dictatorship rule. Thus, $\sigma ^{y}$ is OE. 
\textbf{Q.E.D.}\newline

There may exist some priority list $\alpha $ such that the ETE reassignment
of the result of serial dictatorship rule with $\alpha $ is not OE. To show
this fact, we revisit Example 5. If the priority list $\alpha $ satisfies $%
\alpha _{1}\in A_{1}$ and $\alpha _{2}\in A_{2}$ (or $\alpha _{1}\in A_{2}$
and $\alpha _{2}\in A_{1}$), then the result of the rule with $\alpha $ is $%
y $ and thus the ETE reassignment of $\sigma ^{y}$ is not OE. Otherwise;
that is, if the priority list $\alpha $ satisfies either $\alpha _{1},\alpha
_{2}\in A_{1}$ or $\alpha _{1},\alpha _{2}\in A_{2}$, then the ETE
reassignment of the result of the rule with $\alpha $ is OE. This
observation suggests that the choice of the priority list is crucial for
preserving ordinal efficiency under ETE reassignment.

We consider the following specific priority lists. A priority list $\alpha $
is said to satisfy \textbf{consecutive equals} if, for any$\ a,b\in A$ with $%
a=\alpha _{i}$ and $b=\alpha _{j}$ are equals and $j>i$ imply that all
agents $\alpha _{i},\alpha _{i+1},\cdots ,\alpha _{j}$ are also equals. We
have the following result.

\begin{theorem}
Let $y$ be the result of the serial dictatorship rule with a priority list
that satisfies consecutive equals. If $Y$ satisfies the general upper bounds
constraint, then $\epsilon \left( \sigma ^{y}\right) $ is OE.
\end{theorem}

\textbf{Proof.} Let $\alpha $ be a priority list that satisfies consecutive
equals. We index the groups $A_{1},\cdots ,A_{N}$ along $\alpha $ so that,
whenever $\alpha _{i}$ $\in A_{n}$ and $\alpha _{j}\in A_{m}$, $n<m$ implies 
$i<j$; in particular, $\alpha _{1}\in A_{1}$ and $\alpha _{\left\vert
A\right\vert }\in A_{N}$. Let $y$ be the outcome of the serial dictatorship
rule with $\alpha $. Moreover, let $\sigma ^{\prime }$ be an arbitrary
assignment such that for all $a$, $\mathbf{x}\left( \epsilon \left( \sigma
^{y}\right) \right) _{a}$ is first-order stochastically dominated by $%
\mathbf{x}\left( \sigma ^{\prime }\right) _{a}$. To prove this theorem, it
suffices to show that for all $a\in A$, $\mathbf{x}\left( \epsilon \left(
\sigma ^{y}\right) \right) _{a}=\mathbf{x}\left( \sigma ^{\prime }\right)
_{a}$.

We first establish the following fact.

\begin{claim}
Fix $n\in \left\{ 0,1,\cdots ,N-1\right\} $ and $y^{\prime }\in \mathrm{Supp}%
\left( \sigma ^{\prime }\right) $. Suppose for each $l=1,\cdots ,n$, there
exists a bijection $\pi _{l}:A_{l}\rightarrow A_{l}$ such that $%
y_{a}^{\prime }=y_{\pi _{l}\left( a\right) }$ for all $a\in A_{l}$. Then,
there exists a bijection $\pi _{n+1}:A_{n+1}\rightarrow A_{n+1}$ such that $%
y_{a}^{\prime }=y_{\pi _{n+1}\left( a\right) },$ and $\mathbf{x}\left(
\sigma ^{\prime }\right) _{a}=\mathbf{x}\left( \epsilon \left( \sigma
^{y}\right) \right) _{a}$ for all $a\in A_{n+1}$.
\end{claim}

\textbf{Proof of Claim 2.} Since $\alpha $ satisfies consecutive equals, we
can write $A_{n+1}=\left\{ \alpha _{i},\cdots ,\alpha _{i+j}\right\} $ for
some $j\geq 0$. Since the agents in $A_{n+1}$ have the same preference order
by Assumption 1 and $Y$ satisfies the general upper bounds constraint, we
have 
\begin{equation*}
y_{\alpha _{i}}\succsim _{a}y_{\alpha _{i+1}}\succsim _{a}\cdots \succsim
_{a}y_{\alpha _{i+j}}
\end{equation*}%
for all $a\in A_{n+1}$. Let $x_{1}=y_{\alpha _{i}}$. We first show that no
agent in $A_{n+1}$ can receive an assignment strictly preferred to $x_{1}$
under $y^{\prime }$.

Let $x$ be an arbitrary individual assignment such that $x\succ _{\alpha
_{i}}x_{1}$. First, define $\bar{y}$ by 
\begin{eqnarray*}
\bar{y}_{a} &=&y_{a}\text{ for all }a\in A_{1}\cup \cdots \cup A_{n}, \\
\bar{y}_{a} &=&\mathbf{0}\text{ for all }a\in \left\{ \alpha _{i+1},\cdots
,\alpha _{i+j}\right\} \cup A_{n+2}\cup \cdots \cup A_{N}, \\
\bar{y}_{\alpha _{i}} &=&x.
\end{eqnarray*}
Then, $\bar{y}$ is infeasible.

Next, let $\bar{y}^{\prime }$ be such that $\bar{y}_{a}^{\prime
}=y_{a}^{\prime }$ for all $a\in \left( A_{1}\cup \cdots \cup A_{N}\right)
\setminus \left\{ \alpha _{i}\right\} $ and $\bar{y}_{\alpha _{i}}^{\prime
}=x$. Since there exists a bijection $\pi _{l}:A_{l}\rightarrow A_{l}$ such
that $y_{a}^{\prime }=y_{\pi _{l}\left( a\right) }$ for all $a\in A_{l}$,
and since the agents within each $A_{l}$ are equals for each $l=1,\cdots ,n$%
, $\bar{y}^{\prime }$ is also infeasible (by Assumption 2). Hence $%
x_{1}\succsim _{a}y_{\alpha _{i}}^{\prime }$ for all $a\in A_{n+1}$.

Suppose $y_{\alpha _{i}}=\cdots =y_{\alpha _{i+k}}=x_{1}$ where $k=0,\cdots
,j$; that is, for $y$, there are $k+1$ agents in $A_{n+1}$ who obtain $x_{1}$%
. Then, by the construction of the ETE reassignment, 
\begin{equation*}
\Pr \left( x_{1};\mathbf{x}\left( \epsilon \left( \sigma ^{y}\right) \right)
_{a}\right) =\frac{k+1}{j+1}
\end{equation*}%
for all $a\in A_{n+1}$. Since $\mathbf{x}\left( \epsilon \left( \sigma
^{y}\right) \right) _{a}$ is first-order stochastically dominated\textbf{\ }%
for all $a\in A_{n+1}$ by $\mathbf{x}\left( \sigma ^{\prime }\right) _{a}$,
and since $x_{1}\succsim _{a}y_{a^{\prime }}^{\prime }$ for all $a,a^{\prime
}\in A_{n+1}$, it follows that 
\begin{equation}
\Pr \left( x_{1};\mathbf{x}\left( \sigma ^{\prime }\right) _{a}\right) \geq 
\frac{k+1}{j+1}=\Pr \left( x_{1};\mathbf{x}\left( \epsilon \left( \sigma
^{y}\right) \right) _{a}\right)  \label{y}
\end{equation}%
for all $a\in A_{n+1}$.

For any $y^{\prime }\in \mathrm{Supp}\left( \sigma ^{\prime }\right) ,$
there are at most $k+1$ agents in $A_{n+1}$ who obtain $x_{1}$. If $k=j$,
this is immediate, since $A_{n+1}$ contains exactly $j+1=k+1$ agents. Thus,
suppose $k<j$. Then, 
\begin{equation*}
x_{1}=y_{\alpha _{i}}=\cdots =y_{\alpha _{i+k}}\succ _{a}y_{\alpha _{i+k+1}}%
\text{,}
\end{equation*}%
assigning $x_{1}$ to agent $\alpha _{i+k+1}$ yields a strictly better
assignment for them. By the construction of the serial dictatorship rule, if
we let $\bar{y}$ be an assignment such that 
\begin{equation*}
\bar{y}_{a}=y_{a}\text{ for all }a\in A_{1}\cup \cdots \cup A_{n}\cup
\left\{ \alpha _{i},\cdots ,\alpha _{i+k}\right\}
\end{equation*}%
and $\bar{y}_{\alpha _{i+k+1}}=x_{1}$. By the definition of the serial
dictatorship rule, $\bar{y}$ is infeasible. Since for each $l=1,\cdots ,n,$
there exists a bijection $\pi _{l}:A_{l}\rightarrow A_{l}$ such that $%
y_{a}^{\prime }=y_{\pi _{l}\left( a\right) }$ for all $a\in A_{l}$, and
since $Y$ satisfies the general upper bounds constraint, any assignment that
gives $x_{1}$ to $k+2$ or more agents in $A_{n+1}$ is also infeasible. Hence
for any $y^{\prime }\in \mathrm{Supp}\left( \sigma ^{\prime }\right) ,$
there are at most $k+1$ agents in $A_{n+1}$ who obtain $x_{1}$.

Therefore, 
\begin{equation*}
\sum_{a\in A_{n+1}}\Pr \left( x_{1};\mathbf{x}\left( \sigma ^{\prime
}\right) _{a}\right) \leq k+1\text{.}
\end{equation*}%
On the other hand, as shown above, for every $a\in A_{n+1}$, (\ref{y}) is
satisfied. Therefore 
\begin{equation*}
\Pr \left( x_{1};\mathbf{x}\left( \sigma ^{\prime }\right) _{a}\right) =%
\frac{k+1}{j+1}=\Pr \left( x_{1};\mathbf{x}\left( \epsilon \left( \sigma
^{y}\right) \right) _{a}\right)
\end{equation*}%
for every $a\in A_{n+1}$.

Moreover, for each $y^{\prime }\in \mathrm{Supp}\left( \sigma ^{\prime
}\right) ,$ at most $k+1$ agents in $A_{n+1}$ receive $x_{1}$. Since the
expected number of such agents is equal to $k+1$, for every $y^{\prime }\in 
\mathrm{Supp}\left( \sigma ^{\prime }\right) $, exactly $k+1$ in $A_{n+1}$
receive $x_{1},$ as under $y$.

Next, let $y_{\alpha _{i+k+1}}=x_{2}$, which is the second best pure
assignment for each $a\in A_{n+1}$ under $y$. Applying the preceding
argument to $x_{2}$, while using the results already established for $x_{1}$%
, we obtain 
\begin{equation*}
\Pr \left( x_{2};\mathbf{x}\left( \sigma ^{\prime }\right) _{a}\right) =\Pr
\left( x_{2};\mathbf{x}\left( \epsilon \left( \sigma ^{y}\right) \right)
_{a}\right)
\end{equation*}%
for all $a\in A_{n+1}$. Moreover, under each $y^{\prime }\in \mathrm{Supp}%
\left( \sigma ^{\prime }\right) ,$ the number of agents in $A_{n+1}$ who
receive $x_{2}$ is the same as under $y$.

Iterating this argument over all distinct individual assignments received by
agents in $A_{n+1}$ under $y$, we conclude that $\mathbf{x}\left( \sigma
^{\prime }\right) _{a}=\mathbf{x}\left( \epsilon \left( \sigma ^{y}\right)
\right) _{a}$ for all $a\in A_{n+1}$ and that there exists a bijection 
\begin{equation*}
\pi _{n+1}^{y^{\prime }}:A_{n+1}\rightarrow A_{n+1}
\end{equation*}%
such that $y_{a}^{\prime }=y_{\pi _{n+1}\left( a\right) }$ for all $%
y^{\prime }\in \mathrm{Supp}\left( \sigma ^{\prime }\right) $. \textbf{Q.E.D.%
}\newline

Applying Claim 2 successively for $n=0,1,\cdots ,N-1$, we obtain the
following result. For every $n=1,\cdots ,N$ and every $y^{\prime }\in 
\mathrm{Supp}\left( \sigma ^{\prime }\right) $, there exists a permutation $%
\pi _{n}$ of $A_{n}$ such that $y_{a}^{\prime }=y_{\pi _{n}\left( a\right) }$
for all $a\in A_{n}$. Moreover, $\mathbf{x}\left( \sigma ^{\prime }\right)
_{a}=\mathbf{x}\left( \epsilon \left( \sigma ^{y}\right) \right) _{a}$ for
all $a\in A_{n}$.

Since $A_{1},\cdots ,A_{N}$ form a partition of $A$, it follows that $%
\mathbf{x}\left( \sigma ^{\prime }\right) _{a}=\mathbf{x}\left( \epsilon
\left( \sigma ^{y}\right) \right) _{a}$ for all $a\in A$. Therefore, any
assignment $\sigma ^{\prime }$ that weakly ordinally dominates $\epsilon
\left( \sigma ^{y}\right) $ gives every agent exactly the same distribution
as $\epsilon \left( \sigma ^{y}\right) $. Hence, the dominance cannot be
strict for any agent. Thus, $\sigma ^{\prime }$ cannot strictly ordinally
dominate $\epsilon \left( \sigma ^{y}\right) $, and thus $\epsilon \left(
\sigma ^{y}\right) $ is OE. \textbf{Q.E.D.}\newline

By Theorem 2, the following computationally efficient method derives an
assignment that satisfies both OE and ETE. First, we construct a priority
list that satisfies consecutive equals. Second, we derive an assignment by
using the serial dictatorship rule with the priority list constructed in the
first step. Third, we derive the ETE reassignment of the assignment
constructed in the second step.

Note that any of the results of this method may not be RE. To show this, we
provide the following example.

\subsubsection*{Example 6}

Let 
\begin{gather*}
\succ _{a_{1}}:o_{1},o_{3},o_{2},o_{4}, \\
\succ _{a_{2}}:o_{2},o_{1},o_{3},o_{4}, \\
\succ _{a_{3}}:o_{1},o_{2},o_{3},o_{4}, \\
\succ _{a_{4}}:o_{1},o_{2},o_{3},o_{4}.
\end{gather*}%
Suppose that $a_{3}$ and $a_{4}$ are equals. There are $12$ priority lists
satisfying consecutive equals. Among them, since the outcome of our method
does not depend on the priority difference between $a_{3}$ and $a_{4}$, we
only consider the six priority lists in which $a_{3}$ is ranked higher than $%
a_{4}$. To be precise, the following priority lists are summarized as the
following table meaning that $\alpha _{1}^{1}=a_{1},$ $\alpha
_{2}^{1}=a_{2}, $ $\alpha _{3}^{1}=a_{3},$ and $\alpha _{4}^{1}=a_{4},$ ... .

\begin{center}
$%
\begin{array}{ccccc}
\alpha ^{1}: & a_{1} & a_{2} & a_{3} & a_{4} \\ 
\alpha ^{2}: & a_{1} & a_{3} & a_{4} & a_{2} \\ 
\alpha ^{3}: & a_{2} & a_{1} & a_{3} & a_{4} \\ 
\alpha ^{4}: & a_{2} & a_{3} & a_{4} & a_{1} \\ 
\alpha ^{5}: & a_{3} & a_{4} & a_{1} & a_{2} \\ 
\alpha ^{6}: & a_{3} & a_{4} & a_{2} & a_{1}%
\end{array}%
$
\end{center}

Then, let $y^{i}$ be the result of the serial dictatorship rule with
priority list $\alpha ^{i}$ such that%
\begin{eqnarray*}
y^{1} &=&y^{3}=\left( 
\begin{array}{cccc}
1 & 0 & 0 & 0 \\ 
0 & 1 & 0 & 0 \\ 
0 & 0 & 1 & 0 \\ 
0 & 0 & 0 & 1%
\end{array}%
\right) , \\
y^{2} &=&\left( 
\begin{array}{cccc}
1 & 0 & 0 & 0 \\ 
0 & 0 & 0 & 1 \\ 
0 & 1 & 0 & 0 \\ 
0 & 0 & 1 & 0%
\end{array}%
\right) ,y^{4}=\left( 
\begin{array}{cccc}
0 & 0 & 0 & 1 \\ 
0 & 1 & 0 & 0 \\ 
1 & 0 & 0 & 0 \\ 
0 & 0 & 1 & 0%
\end{array}%
\right) , \\
y^{5} &=&\left( 
\begin{array}{cccc}
0 & 0 & 1 & 0 \\ 
0 & 0 & 0 & 1 \\ 
1 & 0 & 0 & 0 \\ 
0 & 1 & 0 & 0%
\end{array}%
\right) ,y^{6}=\left( 
\begin{array}{cccc}
0 & 0 & 0 & 1 \\ 
0 & 0 & 1 & 0 \\ 
1 & 0 & 0 & 0 \\ 
0 & 1 & 0 & 0%
\end{array}%
\right) .
\end{eqnarray*}%
Thus, if $y^{i}$ is the result of our method above with priority list $%
\alpha ^{i}$, then 
\begin{eqnarray*}
R\left( \sigma ^{y^{1}}\right) &=&R\left( \sigma ^{y^{3}}\right) =R\left(
\sigma ^{y^{4}}\right) =R\left( \sigma ^{y^{5}}\right) =9, \\
R\left( \sigma ^{y^{2}}\right) &=&R\left( \sigma ^{y^{6}}\right) =10.
\end{eqnarray*}%
However, if%
\begin{equation*}
y=\left( 
\begin{array}{cccc}
0 & 0 & 1 & 0 \\ 
0 & 1 & 0 & 0 \\ 
1 & 0 & 0 & 0 \\ 
0 & 0 & 0 & 1%
\end{array}%
\right) ,
\end{equation*}%
then $R\left( \sigma ^{y}\right) =8$. Note that $y$ is achieved by the
serial dictatorship rule with $\alpha $ such that $\alpha _{1}=a_{3},$ $%
\alpha _{2}=a_{1},$ $\alpha _{3}=a_{2},$ and $\alpha _{4}=a_{4}$, which does
not satisfy consecutive equals. Therefore, none of the assignments $\sigma
^{y^{i}}$ for $i=1,\cdots ,6$ is rank-minimizing efficient.\newline

\section{Strategic implications}

Let $f\left( \bar{\succ}\right) $ be a mechanism when agents report $\bar{%
\succ}=\left( \bar{\succ}_{a}\right) _{a\in A}$. Let $\succ _{a}$ be the
true preference of $a$ and $\succ =\left( \succ _{a}\right) _{a\in A}$.
Specifically, $\bar{\succ}=\succ $ means that all agents report their true
preference orders (i.e., truth-telling). For simplicity, we assume that $a$
and $b$ are equals under $\bar{\succ}$ if $\bar{\succ}_{a}=\bar{\succ}_{b}$.
Thus, for given $\bar{\succ}$, let $A_{1}\left( \bar{\succ}\right) ,\cdots
,A_{N}\left( \bar{\succ}\right) $ be a partition of $A$ induced by this
equivalence relation, where $N\leq \left\vert A\right\vert $.

We consider the following simple mechanism $f$. First, for given $\bar{\succ}
$, a pure assignment $y$ is determined by the serial dictatorship rule for
some priority list. Second, we derive the ETE reassignment based on $\bar{%
\succ}$ of $\sigma ^{y}$. Here, the priority list for a preference profile
is allowed to be different from that for another preference profile.
Moreover, we do not require that any priority list satisfy consecutive
equals.

We show that this mechanism is not strategy-proof, even in a weak sense.
That is, for some $\succ $, some $a\in A$, and some $\succ _{a}^{\prime },$ $%
\mathbf{x}\left( f\left( \succ \right) \right) _{a}$ is first-order
stochastically dominated for $a$ by $\mathbf{x}\left( f\left( \succ
_{a}^{\prime },\succ _{-a}\right) \right) _{a}$. Hereafter, if there is some
preference order $\succ _{a}^{\prime }$ such that $\mathbf{x}\left( f\left(
\succ \right) \right) _{a}$ is first-order stochastically dominated by $%
\mathbf{x}\left( f\left( \succ _{a}^{\prime },\succ _{-a}\right) \right)
_{a} $ for $a,$ then we simply say that $a$ has an incentive to manipulate
its preference order.

\subsubsection*{Example 7}

In this example, the agents with identical preference orders are considered
equals. Let $A=\left\{ a_{1},a_{2},a_{3}\right\} $ and $O=\left\{
o_{1},o_{2},o_{3}\right\} $. We consider the unit-demand case and moreover,
the following three preference orders. 
\begin{eqnarray*}
\theta &:&\text{ }o_{1},o_{2},o_{3}. \\
\theta ^{\prime } &:&\text{ }o_{2},o_{1},o_{3}. \\
\theta ^{\prime \prime } &:&\text{ }o_{2},o_{3},o_{1}.
\end{eqnarray*}%
Since there are three agents, there are six possible priority lists
summarized as the following table.

\begin{center}
$%
\begin{array}{cccc}
\alpha ^{1}: & a_{1} & a_{2} & a_{3} \\ 
\alpha ^{2}: & a_{1} & a_{3} & a_{2} \\ 
\alpha ^{3}: & a_{2} & a_{1} & a_{3} \\ 
\alpha ^{4}: & a_{2} & a_{3} & a_{1} \\ 
\alpha ^{5}: & a_{3} & a_{1} & a_{2} \\ 
\alpha ^{6}: & a_{3} & a_{2} & a_{1}%
\end{array}%
$
\end{center}

We consider the case $\left( \bar{\succ}_{a_{1}},\bar{\succ}_{a_{2}},\bar{%
\succ}_{a_{3}}\right) =\left( \theta ,\theta ^{\prime },\theta ^{\prime
\prime }\right) $. If $\alpha ^{2}$, $\alpha ^{5},$ or $\alpha ^{6}$ is
used, then $o_{2}$ is assigned to $a_{3}$ with probability $1$. If $\alpha
^{1}$, $\alpha ^{3},$ or $\alpha ^{4}$ is used, then $o_{2}$ is assigned to $%
a_{2}$ with probability $1$.

First, we assume that $\alpha ^{2}$, $\alpha ^{5},$ or $\alpha ^{6},$ is
used when $\left( \bar{\succ}_{a_{1}},\bar{\succ}_{a_{2}},\bar{\succ}%
_{a_{3}}\right) =\left( \theta ,\theta ^{\prime },\theta ^{\prime \prime
}\right) $. Then, $f\left( \theta ,\theta ^{\prime },\theta ^{\prime \prime
}\right) _{a_{3}o_{2}}=1$. Suppose that the true preferences of the agents
are $\left( \succ _{a_{1}},\succ _{a_{2}},\succ _{a_{3}}\right) =\left(
\theta ,\theta ^{\prime },\theta ^{\prime }\right) $. Then, since $a_{2}$
and $a_{3}$ are equals under the true preferences and $f\left( \theta
,\theta ^{\prime },\theta ^{\prime }\right) $ should satisfy ETE, 
\begin{equation*}
\Pr \left( o_{2};\mathbf{x}\left( f\left( \theta ,\theta ^{\prime },\theta
^{\prime }\right) \right) _{a_{3}}\right) \leq 1/2.
\end{equation*}%
Therefore, $a_{3}$ has an incentive to manipulate its preference order as $%
\theta ^{\prime \prime }$.

Second, we assume that $\alpha ^{1}$, $\alpha ^{3},$ or $\alpha ^{4}$ is
used when $\left( \bar{\succ}_{a_{1}},\bar{\succ}_{a_{2}},\bar{\succ}%
_{a_{3}}\right) =\left( \theta ,\theta ^{\prime },\theta ^{\prime \prime
}\right) $. Then, $f\left( \theta ,\theta ^{\prime },\theta ^{\prime \prime
}\right) _{a_{2}o_{2}}=1$. Suppose that the true preferences of the agents
are $\left( \succ _{a_{1}},\succ _{a_{2}},\succ _{a_{3}}\right) =\left(
\theta ,\theta ^{\prime \prime },\theta ^{\prime \prime }\right) $. Then,
since $a_{2}$ and $a_{3}$ are equals under the true preferences and $f\left(
\theta ,\theta ^{\prime \prime },\theta ^{\prime \prime }\right) $ should be
satisfied ETE, 
\begin{equation*}
\Pr \left( o_{2};\mathbf{x}\left( f\left( \theta ,\theta ^{\prime \prime
},\theta ^{\prime \prime }\right) \right) _{a_{2}}\right) \leq 1/2.
\end{equation*}%
Therefore, $a_{2}$ has an incentive to manipulate its preference order as $%
\theta ^{\prime \prime }$.\newline

The serial dictatorship rule is strategy-proof. However, the mechanism
obtained by naively applying our construction is not even weakly
strategy-proof. To satisfy ETE, the mechanism applies the ETE reassignment
procedure to the outcome of serial dictatorship when some agents report
identical preferences. As Example 7 shows, an agent may benefit by reporting
a preference order different from that reported by an otherwise identical
agent, thereby avoiding being treated as an equal. Thus, the
strategy-proofness of serial dictatorship is not preserved under this
construction. This contrasts with random serial dictatorship, which is
strategy-proof, and probabilistic serial, which is weakly strategy-proof
with respect to stochastic dominance in the standard unit-demand model with
strict preferences.

\section{Concluding Remarks}

In this paper, we discuss assignments that satisfy both efficiency and ETE
by employing the ETE reassignment procedure. We extend the notion of ETE to
accommodate policy objectives such as affirmative action. In our
formulation, characteristics other than preferences---such as gender, race,
or economic disadvantage---may be incorporated into the definition of
equality. However, if too many characteristics are taken into account, few,
if any, agents will be regarded as equals, and ETE will be satisfied only
vacuously. We therefore conclude by discussing the problems associated with
defining equality too narrowly and argue that the relevant characteristics
should be selected so as not to deprive ETE of its substantive content.

First, except where differential treatment is justified by policy objectives
such as affirmative action, basing assignment decisions on inherent personal
characteristics may be normatively objectionable and, depending on the
jurisdiction and context, legally prohibited.

Second, concerns also arise when agents are distinguished according to
characteristics shaped by their economic choices. For example, in the
Japanese day-care matching market, when parents with similar employment
conditions apply to the same day-care center, ties may be broken on the
basis of factors such as household income or length of residence in the
municipality.\footnote{%
Takenami (2025) notes that, in Tama City, when parents receive the same
score based on their working hours, ties are broken using factors such as
length of residence and household income.}

Such rules may distort important economic decisions. For instance,
income-based priority may weaken incentives to increase labor supply, while
residence-based priority may discourage households from relocating.
Accordingly, using characteristics that agents can strategically influence
to differentiate among them raises significant concerns.

Academic performance provides another example. Assigning students on the
basis of entrance examinations may be justified as a means of promoting
academic achievement. At the same time, excessive reliance on examination
performance may intensify educational competition and impose substantial
costs on students and their families. It is therefore neither feasible nor
necessarily desirable to differentiate among agents on the basis of every
observable personal characteristic. Some agents should consequently remain
classified as equals.

Thus, ETE should not be regarded as a vacuous requirement that can be
avoided merely by refining the classification of agents. Once the
characteristics legitimately relevant to assignment decisions are
appropriately restricted, nontrivial groups of equals remain. ETE then
becomes a substantive fairness requirement, and computing assignments that
satisfy it---using methods such as those developed in this paper---becomes
important.

\section*{References}

\begin{description}
\item Abdulkadiro\u{g}lu, A., S\"{o}nmez, T. 2003. School choice: A
mechanism design approach.\ American Economic Review 93(3), 729--747.

\item Andersson, T., Ehlers, L. 2020. Assigning Refugees to Landlords in
Sweden: Efficient, Stable, and Maximum Matchings. The Scandinavian Journal
of Economics 22(3), 937-965.

\item Ayg\"{u}n, O., Bo, I. 2021. College admission with multidimensional
privileges: The Brazilian affirmative action case. American Economic
Journal: Microeconomics 13(3), 1--28.

\item Ayg\"{u}n, O., Turhan, B. 2017. Large Scale Affirmative Action in
School Choice: Admissions to ITTs in India. American Economic Review P\&P
107(5), 210-213

\item Ayg\"{u}n, O., Turhan, B. 2020. Dynamic Reserves in Matching Markets.
Journal of Economic Theory 188, 105069

\item Balbuzanov, I. 2022. Constrained random matching. Journal of Economic
Theory 203, 105472

\item Basteck, C., Ehlers, L. 2025. On (constrained) efficiency of
strategy-proof random assignment. Econometrica 93(2), 569--595

\item Bogomolnaia, A., Moulin, H. 2001. A New Solution to the Random
Assignment Problem. Journal of Economic Theory 100, 295-328.

\item Budish, E., Cantillon, E. 2012. The multi-unit assignment problem:
Theory and evidence from course allocation at harvard. American Economic
Review 102(5), 2237-2271.

\item Budish, E., Che, Y.-K., Kojima, F., Milgrom, P. 2013. Designing Random
Allocation Mechanisms: Theory and Applications. American Economic Review
103(2), 585--623.

\item Erdil, A. 2014. Strategy-proof stochastic assignment. Journal of
Economic Theory 151, 146-162.

\item Featherstone, C. 2020. Rank efficiency: Modeling a common policymaker
objective. University of Pennsylvania. Unpublished paper.

\item Fehr, E., Charness, G. 2025. Social Preferences: Fundamental
Characteristics and Economic Consequences. Journal of Economic Literature
63(2), 440--514.

\item Feizi, M. 2024. Notions of Rank Efficiency for the Random Assignment
Problem. Journal of Public Economic Theory 26(6), e70008

\item Han, X. 2024. A theory of fair random allocation under priorities.
Theoretical Economics 19, 1185--1221.

\item Hylland, A., Zeckhauser, R. 1979. The efficient allocation of
individuals to positions. Journal of Political Economy 87, 293--314.

\item Imamura, K., Kawase, Y. 2025. Efficient and strategy-proof mechanism
under general constraints. Theoretical Economics 20, 481--509.

\item Kesten, O., Kurino, M., Nesterov, A.S. 2017. Efficient lottery design.
Social Choice and Welfare 48, 31--57.

\item Kesten, O., \"{U}nver, M. 2015. A theory of school choice lotteries.
Theoretical Economics 10, 543--595.

\item Kojima, F., 2009. Random assignment of multiple indivisible objects.
Mathematical Social Sciences 57, 134--142.

\item Kornbluth, D., Kushnir, A., Nguyen, T., Vohra, R. 2025. Comment on
\textquotedblleft Assignment problems with
complementarities\textquotedblright\ [J. Econ. Theory 165 (2016) 209-241].
Journal of Economic Theory, 106130

\item Korte, B., Vygen, J. 2005. Combinatorial Optimization: Theory and
Algorithms (3rd ed.). Springer

\item Kurata, R., Hamada, N., Iwasaki, A., Yokoo, M. 2017. Controlled school
choice with soft bounds and overlapping types. Journal of Artificial
Intelligence Research 58, 153--184.

\item Moulin, H. 2004. Fair division and collective welfare. MIT press

\item Nikzad, A. 2022. Rank-optimal assignments in uniform markets.
Theoretical Economics 17, 25--55.

\item Nguyen, T., Peivandi, A., Vohra, R. 2016. Assignment problems with
complementarities. Journal of Economic Theory 165, 209-241

\item Okumura, Y. 2019. School Choice with General Constraints: A Market
Design Approach for the Nursery School Waiting List Problem in Japan.
Japanese Economic Review 70(4), 497-516

\item Okumura, Y. 2026a. Strategic Analysis of Fair Rank-Minimizing
Mechanisms with Agent Refusal Option. Journal of Mathematical Economics 124,
103250

\item Okumura, Y. 2026b. A Simple Method for School Choice Lotteries.
arXiv:2605.06721

\item Ortega, J., Klein, T. 2023. The cost of strategy-proofness in school
choice. Games and Economic Behavior 141, 515--528

\item Roth, A.E., Sotomayor, M.A.O. 1990. Two-Sided Matching A Study in
Game-Theoretic Modeling and Analysis. Cambridge University Press

\item S\"{o}nmez, T., \"{U}nver, M. 2010. Course Bidding at Business
Schools. International Economic Review 51(1), 99-123.

\item S\"{o}nmez, T., Yenmez, M. B. 2022. Affirmative action in India via
vertical, horizontal, and overlapping reservations. Econometrica 90(3),
1143--1176.

\item Takenami, Y. 2025. Making nursery school admissions fair so that
honest parents don't lose out. (written in Japanese) in AI and Economics,
written by Moriwaki, D., Takenami, Y., Tomita, Y., Yamada, N. 173--203.

\item Thomson, W. 2011. Fair Allocation Rules. Ch. 21 in Handbook of Social
Choice and Welfare, eds. by Arrow, K.-J., Sen, A., Suzumura, K., Vol. 2,
393--506.

\item Troyan, P. 2024. (Non-)obvious manipulability of rank-minimizing
mechanisms. Journal of Mathematical Economics 113, 103015

\item Varian, H.R., 1974. Equity, envy, and efficiency. Journal of Economic
Theory 9(1), 63-91.
\end{description}

\end{document}